%% file: paper.tex
\documentclass{siamltex}
\usepackage{cite}
\usepackage{amsmath,amssymb,amsfonts}
\usepackage{algorithmic,algorithm}
\usepackage{graphicx}
\usepackage{textcomp,multirow,subcaption,listings,comment}
\usepackage{rotating}

\input{preamble.tex}

\graphicspath{{./FIGS/}}

\begin{document}

%

\title{On Parallel Solution of Sparse Triangular Linear Systems in CUDA}
\author{Ruipeng Li\thanks{Center  for Applied  Scientific Computing,
    Lawrence  Livermore National  Laboratory,  P. O.  Box 808,  L-561,
    Livermore,   CA  94551   {(\tt{li50@llnl.gov})}.  This   work  was
    performed under the  auspices of the U.S. Department  of Energy by
    Lawrence    Livermore   National    Laboratory   under    Contract
    DE-AC52-07NA27344 (LLNL-JRNL-739550).}}

\maketitle

\begin{abstract}
The acceleration of sparse matrix computations on modern many-core processors, such as the graphics processing units (GPUs), has been 
recognized and studied over a decade. Significant performance enhancements have been achieved for many sparse matrix computational kernels
such as sparse matrix-vector products and sparse matrix-matrix products.
Solving linear systems with sparse triangular structured  matrices is another important 
sparse  kernel as  demanded by a variety of
scientific and engineering applications such as sparse linear solvers.
However, the development of efficient parallel algorithms in CUDA for solving sparse triangular linear systems  remains a challenging task due to the inherently sequential nature of the computation.
In this paper, we will revisit this problem by reviewing the existing level-scheduling methods and proposing   algorithms with self-scheduling techniques.
Numerical results have indicated that the CUDA implementations of the proposed algorithms can outperform the state-of-the-art solvers in cuSPARSE by a factor 
of up to $2.6$ for structured model problems and general sparse matrices.
\end{abstract}



\section{Introduction} \label{sec:intro}
The emerging many-core architectures can deliver enormous raw processing power in the form of
massive single-instruction-multiple-data (SIMD) parallelism.
The graphics processing unit (GPU) is one of the several available platforms that features a large number of cores.
Traditionally, GPUs were aimed to handle the computations in real-time computer graphics. 
However, they are now increasingly being exploited as general-purpose processors for scientific computations.
The potential of GPUs for sparse matrix computations was recognized in the early 2000s
when GPU programming still required shading languages 
\cite{Bolz03sparsematrix,Kruger:2003:LAO:882262.882363}. Since the advent of
CUDA, the NVIDIA GPUs have drawn much more attention for accelerating sparse matrix computations such as
sparse linear solvers \cite{doi:10.1137/140980260,RENNICH2016140,Sao2014,Wang2009,CLARK20101517,6749152,Gandham20141151,doi:10.1080/17445760802337010,RliSaadGPU,doi:10.1137/15M1026419} and sparse eigensolvers \cite{Anzt:2015:ALM:2872599.2872609,dziekonski_rewienski_sypek_lamecki_mrozowski_2017,AURENTZ2017332}.

Significant performance enhancements have been achieved on GPUs for sparse matrix computational kernels
such as sparse matrix-vector products \cite{BellSpMV,Liu2015179,Liu:2015:CES:2751205.2751209,Choi:2010:MAS:1837853.1693471,Ashari:2014:FSM:2683593.2683679,baskaran2008optimizing} and sparse matrix-matrix products \cite{Liu:2014:EGG:2650283.2650651,6507483,doi:10.1137/110838844},
where the computations are massively parallelizable.
However, the parallel solution of sparse triangular linear systems  remains a challenging task on GPUs
due to its inherently sequential nature.
Compared with the multiplication kernels, the performance reached by  sparse triangular solve kernels is much lower.
Solving sparse triangular systems is demanded by a variety of scientific and engineering applications.
For example, these solves are the essence of the solve phases of sparse direct methods for solving linear systems,
and are also the operations of applying Gauss-Seidel type relaxation methods and
incomplete LU (ILU) factorization type preconditioners in iterative solvers.

This issue actually motivated the development of a class of sparse approximate inverse preconditioners in the 1990s 
as alternatives to ILU preconditioners.
In this type of method, instead of computing the factorizations of the coefficient matrix $A$, 
approximate inverses of $A$ or approximate  factors of $A\inv$ are pursued.
Thus, the preconditioning operations  involve sparse matrix-vector multiplications,
which can often yield much higher throughput than the solves on parallel machines. On the other hand, by and large,
the cost of computing  approximate inverse preconditioners and the number of iterations required for
convergence is higher than those for ILUs, especially for ill-conditioned or indefinite systems.
Consequently, there still remains the need for efficient application of ILU preconditioning in  massively parallel environments. 
The same situation also exists in algebraic multigrid (AMG) methods, where Gauss-Seidel type smoothers often turn out to be more robust and
preferable than their Jacobi type counterparts.
Therefore, in all these scenarios, to avoid a sequential bottleneck, 
efficient parallel algorithms for solving sparse triangular systems are of great demand for  modern many-core processors.

In this paper, we will revisit this problem by reviewing the existing algorithms based on level-scheduling approaches 
and proposing algorithms with self-scheduling techniques.
The implementations of these algorithms in CUDA will be discussed in great detail.
In a nutshell, the idea in the level-scheduling methods is to group the unknowns into different levels such that
the unknowns within the same level are free of dependencies and can be solved simultaneously. Moreover, 
global synchronizations  are typically required  to resolve the dependencies between the levels.
In contrast, more aggressive scheduling schemes are used in the proposed self-scheduling algorithms, where
the computations for an unknown are immediately started as soon as the solutions of  the unknowns
that this unknown depends on
are all available, and
there is no global synchronization required in these algorithms.
It is worth pointing out here that the different scheduling approaches mentioned will not change the degree of parallelism
in the solve, which is
on average determined by the total number of  tasks (unknowns) and the number of levels.
The focus of the parallel algorithms studied in this work is to explore different 
scheduling schemes to resolve the dependencies and schedule the parallel tasks more efficiently by
minimizing the latency between the time when a task is ready to be started and  its actual starting time.

All the parallel algorithms for solving sparse triangular systems discussed in this paper require a setup phase (or called an
analysis phase), where the parallelism in the solve phase will be discovered through the nonzero patterns of the sparse triangular matrices.
Hence, the setup phase should be considered as an extra overhead of the parallel algorithms compared with the naive approaches. 
However, several factors should be taken into account: First, the  setup phases are often not too expensive, 
relative to the cost of the solve phases, to make the use of the parallel algorithms unrealistic. So, in many cases, it still  pays off
to run the parallel algorithms with the setup phases. Second, there  are  many  applications
in  which  multiple right-hand sides   with  the  same  triangular matrix  need to  be  solved. 
For instance, in  iterative methods with ILU preconditioners, triangular solves are required in every iteration.
This is also true for  sparse direct methods when multiple steps of iterative refinements are performed after the solve.
Therefore, in these situations, the overhead in the setup phase may be justified because the cost can be amortized by the multiple solves. 
Third, in a multi-processor environment, for certain applications, the setup phase can be overlapped with other computations once the 
patterns of the triangular matrices are known. For  Gauss-Seidel  relaxation and the ILU factorization without fill-in,
the triangular factors have the same nonzero structures as the lower and upper parts of the original matrix.
For sparse direct methods or  ILU factorizations with level-based fill-in, the setup phase can be started immediately
after the symbolic factorization is done and can be executed in parallel with the numerical factorization stage. 

The remaining of the paper is organized as follows:  
a summary of related work will be given in Section~\ref{sec:relatedwork};
we will review the row-wise and the column-wise forward and backward sweeps for solving triangular systems in Section~\ref{sec:prelim},
which serve as the fundamental frameworks of the parallel algorithms discussed in
Section~\ref{sec:alg};   numerical  results  of  model  problems  and  general  matrices 
will be presented in Section~\ref{sec:num} and 
 we conclude in Section~\ref{sec:conclusion}.
 
\section{Related work} \label{sec:relatedwork}
To the best of our knowledge, the first level-scheduling type algorithms for the parallel solution 
of sparse triangular systems  were
due to Anderson and Saad \cite{Anderson:1989:SST:85085.85106}, and later by Saltz in \cite{doi:10.1137/0911008}, where 
the forward and backward substitutions 
are scheduled into levels or  so-called wavefronts.
The self-scheduling algorithms
for shared-memory parallel  machines were first introduced in \cite{George1986}.
A related body of literature also exists on the algorithms for solving triangular systems on distributed-memory processors, see, e.g., \cite{doi:10.1137/0909037, doi:10.1137/0909038,doi:10.1137/0909032}.
The works in \cite{naumov2011parallel,RliSaadGPU} might be the first efforts on 
implementing the level-scheduling algorithms in CUDA, and 
similar approaches were later adopted in \cite{6337473,7839683}.
More recent work on the synchronization-free algorithm on GPUs can be found in \cite{Liu2016},
and on solving sparse triangular systems with multiple right-hand sides in \cite{CPE:CPE4086}. 
For fast dense triangular solves in CUDA, see \cite{doi:10.1137/12088358X}.

Research has also been done on the impact of matrix reorderings and colorings on
the performance of sparse triangular solves on GPUs \cite{RliSaadGPU,6337473,6267851,naumov2015parallel}. 
In general, the number of levels in the substitutions in the triangular parts of a  matrix 
or its LU factors can be significantly reduced if
the matrix has been relabeled by graph colorings or by certain types of matrix reorderings \cite{RliSaadGPU}.
However, we must point out that the relabeling corresponds to permuting the  matrix or
equivalently performing relaxations in different orders, and actually changes the triangular system to  solve.
Thus, graph colorings and matrix reorderings will not be considered in this paper.

Another line of parallel methods for solving sparse triangular systems is 
based on the ``partitioned inverses'' proposed in \cite{doi:10.1137/0914027, Alvarado1993}, 
where the inverse of the triangular matrix is represented 
as a product of a few sparse factors. Thus, solving the triangular system 
reduces to a few matrix-vector products with triangular matrices.
Related works also include the approach of using iterative methods to approximately
solve sparse triangular systems  for preconditioning purposes \cite{doi:10.1137/140968896, Anzt2015}.
These types of methods are also out of the scope of this work. 
In this paper, we will focus on parallel algorithms for solving sparse triangular systems \emph{exactly}, based on
\emph{forward and backward substitutions}, and their efficient implementations in CUDA.


\section{Preliminaries} \label{sec:prelim}
The sparse matrix computational kernel considered in this paper is the sparse triangular solve
(SpTrSv) of the form 
\begin{equation} \label{eq:trisol}
(L+D) x=f  \quad \mbox{or} \quad  (U+D)x=f,
\end{equation}
where $L\in \RR^{n\times n}$ and $U\in \RR^{n\times n}$ are 
sparse  matrices that are strictly lower and upper triangular, $D$ is a diagonal matrix
that is 
assumed to have no zero entries on the diagonal,
and $f \in \RR^{n}$ is the dense right-hand-side vector.
The triangular matrices are assumed to be stored 
in either the compressed sparse row (CSR) format or the compressed sparse column (CSC) format,
and the diagonal matrix is stored separately in a vector $d$, i.e., $D=\mbox{diag}(d)$.

Throughout the paper, we use the notation $(ia, ja, a)$ for the 3 arrays of a CSR matrix that contain the row pointers, 
the column indices and the numerical values respectively. Similarly, $(ib, jb, b)$ is used for a CSC format matrix.
The two different ways of accessing  matrix elements in the CSR  and the CSC formats give rise to 
two fundamental algorithms of performing  forward and backward substitutions 
for solving  sparse triangular systems,  which will be shown in the next two sections in turn.

\subsection{Row-wise SpTrSv}
Supposing that $x$ is first initialized as $f$, the forward substitution
for solving the lower triangular system $(L+D)x=f$ is as follows, where the lower triangular matrix $L$ is assumed to be in the CSR format.

\begin{algorithmic}[1]
\FOR{$i=1,2,\ldots,n$}
\FOR{$j=ia(i),\ldots,ia(i+1)-1$}
\STATE $x(i) := x(i) - a(j)\times x(ja(j))$
\ENDFOR
\STATE $x(i) := x(i) / d(i)$
\ENDFOR
\end{algorithmic}
 
\subsection{Column-wise SpTrSv} \label{sec:colsolve}
When the matrix $L$ is stored in the CSC format, the forward substitution can be done as follows, 
where the elements in the solution $x$ are updated by the columns of $L$.
The initialization step $x:=f$ should have been done before the sweep.

\begin{algorithmic}[1]
\FOR{$i=1,2,\ldots,n$}
\STATE $x(i) := x(i) / d(i)$
\FOR{$j=ib(i),\ldots,ib(i+1)-1$}
\STATE $x(jb(j)) := x(jb(j)) - b(j)\times x(i)$
\ENDFOR
\ENDFOR
\end{algorithmic}
The  backward substitution for solving  upper triangular
systems is similar, where the outer loop should be performed in the 
reverse order, i.e.,
\textbf{for} {$i=n,n-1,\ldots,1$} \textbf{do}.

\section{Parallel SpTrSv algorithms on shared-memory  machines} \label{sec:alg}
In the forward  sweeps shown in the previous section, 
the outer loops  are executed sequentially, whereas the inner loops 
can be vectorized.
In the row-wise sweep, the inner loop computes dot
products of matrix rows and  $x$, while  in the column-wise sweep,
it involves  AXPY operations with matrix columns and $x$.
However, since the number of the entries involved in the inner 
loops is typically small,
this parallelization is often inefficient and
the potential performance benefit from the vectorization is 
usually  outweighed by its overhead.

\subsection{Level-scheduling algorithms} \label{sec:levlsched}

Better parallelism can be achieved by analyzing the dependencies between  unknowns.
Unknown $x(i)$ can be immediately determined once all the others
involved in
equation $i$ become available.
The dependencies can be analyzed by exploiting
the underlying  directed acyclic graph (DAG)  of the triangular matrix.
We associate the $(i,j)$ entry of the matrix if it is nonzero to
the edge from node $j$ to node $i$ in the DAG, which indicates that the solution of $x(i)$  depends on that of $x(j)$.
The idea is then to group the unknowns into different levels, where the first level consists of 
the nodes in the graph with zero in-degree 
and  nodes in any level should only depend on
those in the lower levels.
Therefore, 
the system can be solved level by level and the unknowns within the same level can be computed simultaneously.
The levels of the unknowns can be easily obtained by exploiting a type of topological sorting
of the DAG, which is  referred to as  \emph{level scheduling}  \cite{Saad-book2}.

Denoting by $lev(i)$ the level of unknown $i$,   the forward substitution for $lev(i)$ can be computed as
follows

\begin{algorithmic}[1]
\FOR{$i=1,2,\ldots,n$}
\STATE $lev\left(i\right)=1+{\max}\left\{lev\left(j\right)\right\}, 
\mbox{ for } j \mbox{ such that } L_{ij} \neq 0$
\ENDFOR
\end{algorithmic}
where the $i$-th row of matrix $L$ is accessed to determine $lev(i)$. 
When it is easier to access the matrix elements in columns, $lev(i)$ can be computed by

\begin{algorithmic}[1]
\FOR{$j=1,2,\ldots,n$}
\STATE $lev\left(i\right) =  \max\{lev(j)+1, lev(i)\},
\mbox{ for } i \mbox{ such that } L_{ij} \neq 0$
\ENDFOR
\end{algorithmic}
where $lev(i)$, for $i=1,\ldots,n$, must be first initialized to zeros before running the for-loop.
The levels in the backward substitution can be computed in the same way
by reversing the order of the above computations.
Also, note that the row-wise  and the column-wise algorithms
discussed in Section~\ref{sec:prelim} share the same level information.

Suppose that  $nlev$ denotes the number of levels,  $jlev$ is an array that lists the 
unknowns in a nondecreasing order of their levels, and array $ilev$ contains
the pointers to  the levels in $jlev$.
The row-wise forward substitutions with level scheduling for CSR format matrices
are presented in Algorithm~\ref{alg:LEVR}, where
the second for-loop  computes all the unknowns in level $m$, so 
it can be done in parallel. 

\begin{algorithm}[ht]
\caption {LEVR: Row-wise forward substitutions with level-scheduling \label{alg:LEVR}}
\begin{algorithmic}[1]
\FOR{$m=1,\ldots,nlev$}
\FOR[in parallel]{$k=ilev(m),\ldots,ilev(m+1)-1$} 
\STATE $i=jlev\left(k\right)$
\FOR{$j=ia\left(i\right),\ldots,ia\left(i+1\right)-1$}
\STATE $x(i) := x(i) - a(j)\times x(ja(j))$
\ENDFOR
\STATE $x(i) := x(i) / d(i)$
\ENDFOR
\ENDFOR
\end{algorithmic}
\end{algorithm}

Likewise, the column-wise
algorithm for CSC format matrices is shown in Algorithm~\ref{alg:LEVC}, where
a remarkable difference 
is the requirement of a critical section around the concurrent updates to $x$ 
in order to avoid memory read and write conflicts.

\begin{algorithm}[htbp]
\caption {LEVC: Column-wise forward substitutions with level-scheduling \label{alg:LEVC}}
\begin{algorithmic}[1]
\FOR{$m=1,\ldots,nlev$}
\FOR[in parallel]{$k=ilev(m),\ldots,ilev(m+1)-1$}
\STATE $i=jlev\left(k\right)$
\STATE $x(i) := x(i) / d(i)$
\FOR{$j=ib\left(i\right),\ldots,ib\left(i+1\right)-1$}
\STATE CRITICAL SECTION ENTRY
\STATE $x(jb(j)) := x(jb(j)) - b(j) \times x(i)$
\STATE CRITICAL SECTION EXIT
\ENDFOR
\ENDFOR
\ENDFOR
\end{algorithmic}
\end{algorithm}

\begin{figure}[ht]
\caption{Level scheduling for the 5-point stencil operator a 2-D regular grid\label{fig:level}}
\centerline{\includegraphics[width=0.6\textwidth]{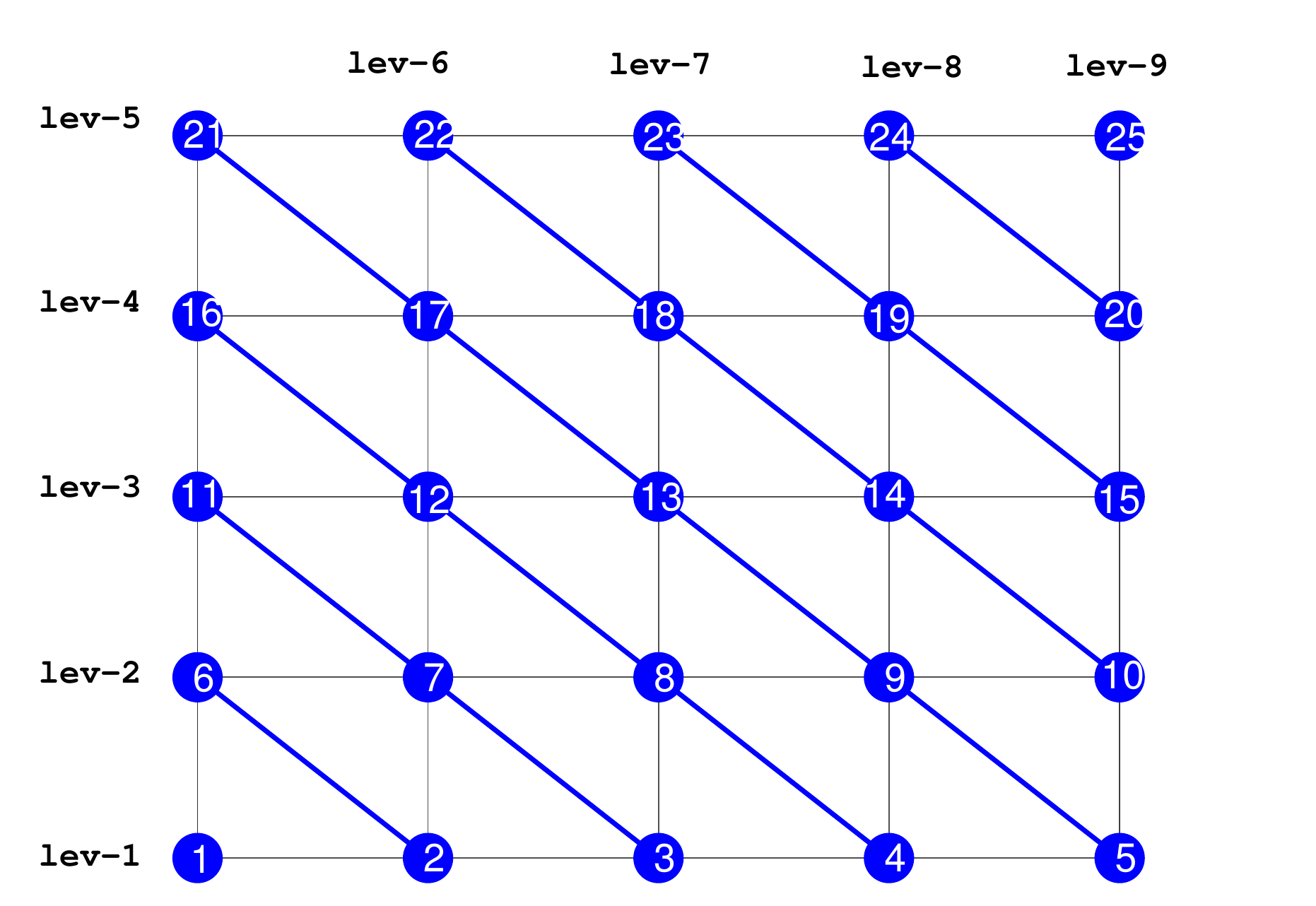}}
\end{figure}

Clearly, in the level-scheduling algorithms, we have $1\leq nlev\leq n$, and thus
the degree of parallelism is $n/nlev$  on average. 
The best case corresponds to the situation where
all the unknowns can be computed simultaneously (i.e., for diagonal matrices where $nlev=1$). 
In the worst case, when $nlev=n$, each unknown will be of a different level,
so that
the entire sweep will become completely sequential.
Therefore, the performance of the level scheduling approaches will
significantly depend on the number of levels.
For a 2-D regular grid of size $n_x \times n_y$,
the number of levels for the lower (or the upper) triangular parts of 5-point operators is given by $nlev=n_x+n_y-1$, 
while for  7-point operators on a 3-D grid of size $n_x \times n_y \times n_z$, 
we have 
$nlev=n_x+n_y+n_z-2$.
In Figure~\ref{fig:level}, we show an example of the levels for 5-point  
operators on a $5\times 5$ grid. 

\subsection{Self-scheduling algorithms}

The level-scheduling algorithms discussed in the previous section have a main drawback of including
a global synchronization point between two levels. It is not only that the 
synchronization itself represents an extra overhead but the synchronization can 
also stall jobs that are ready to go. In the example shown in Figure~\ref{fig:level}, suppose that when dealing with level 5, 
computations on nodes 5, 9 and 13 are finished much earlier than those on  17 and 21. 
In the level-scheduling schemes, the computations on nodes 10 and 14 will not be started
until the entire level 5 finishes although nodes 10 and 14 are free of dependencies immediately after the jobs on nodes 5, 9 and 13 are done.
This issue motivated the development of the  self-scheduling algorithms
that can dynamically schedule the computations on the nodes that are ready and can avoid global synchronizations.

In the self-scheduling algorithms, each node (or equivalently, each unknown)  maintains a counter of the
nodes that it depends on that have not finished. Once the counter reaches
zero, it means that it is ready to solve its unknown. 
A sketch of the row-wise self-scheduling algorithm is given in Algorithm~\ref{alg:SLFR},
where additional auxiliary data for the matrix are required that are the column
pointers $ib$ and the row indices   $jb$ as in the CSC format.
Lines 2--4 make up a busy-waiting loop that iterates until there are no unfinished
dependencies for node $i$. In line 11, the counters of the nodes that
depend on $i$ are decreased by 1, and we remark that line 11 should be put inside
a critical region since the decrements can be executed in parallel from different unknowns. 

\begin{algorithm}[htbp]
\caption {SLFR: Row-wise  forward substitutions with self-scheduling\label{alg:SLFR}}
\begin{algorithmic}[1]
\FOR[rows in parallel]{$i=1,2,\ldots,n$}
\WHILE{$count(i) \neq 0$ }
\STATE \COMMENT{do nothing}
\ENDWHILE
\FOR{$j=ia(i),\ldots,ia(i+1)-1$}
\STATE $x(i) := x(i) - a(j)\times x(ja(j))$
\ENDFOR
\STATE $x(i) := x(i) / d(i)$
\FOR{$j=ib(i),\ldots,ib(i+1)-1$}
\STATE CRITICAL SECTION ENTRY
\STATE $count(jb(j)) := count(jb(j)) - 1$
\STATE CRITICAL SECTION EXIT
\ENDFOR
\ENDFOR
\end{algorithmic}
\end{algorithm}

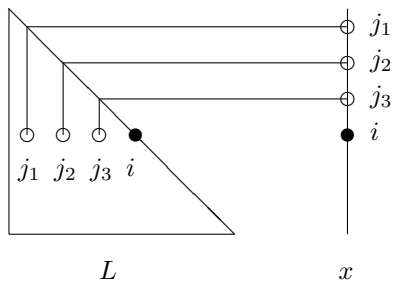
\begin{figure}[htb]
\caption{An illustration of the operations for solving  $x(i)$ in the row-wise forward substitution \label{fig:triag}}
\setlength{\unitlength}{6cm}
\begin{picture}(1.5, 0.55)(-0.7,-0.05)
\put(0,0){\line(0,1){0.5}}
\put(0,0){\line(1,0){0.5}}
\put(0,0.5){\line(1,-1){0.5}}
\put(0.75,0){\line(0,1){0.5}}

\put(0.04,0.22){\circle{0.03}}
\put(0.12,0.22){\circle{0.03}}
\put(0.20,0.22){\circle{0.03}}
\put(0.28,0.22){\circle*{0.03}}

\put(0.04,0.22){\line(0,1){0.24}}
\put(0.75,0.46){\line(-1,0){0.71}}
\put(0.12,0.22){\line(0,1){0.16}}
\put(0.75,0.38){\line(-1,0){0.63}}
\put(0.20,0.22){\line(0,1){0.08}}
\put(0.75,0.30){\line(-1,0){0.55}}

\put(0.75,0.46){\circle{0.03}}
\put(0.75,0.38){\circle{0.03}}
\put(0.75,0.30){\circle{0.03}}
\put(0.75,0.22){\circle*{0.03}}

\put(0.02,0.13){$j_1$}
\put(0.10,0.13){$j_2$}
\put(0.18,0.13){$j_3$}
\put(0.26,0.13){$i$}

\put(0.8,0.45){$j_1$}
\put(0.8,0.37){$j_2$}
\put(0.8,0.29){$j_3$}
\put(0.8,0.21){$i$}

\put(0.2,-0.1){$L$}
\put(0.73,-0.1){$x$}
\end{picture}
\end{figure}

The column-wise self-scheduling algorithm for forward substitutions is presented in 
Algorithm~\ref{alg:SLFC}, which is based on the same counter-based scheme, and
the input counters  are identical to those used in the row-wise algorithm.
Comparing with Algorithm~\ref{alg:SLFR}, first this algorithm does not require additional inputs for the matrix  other than the arrays 
in the CSC format.
Second, Algorithm~\ref{alg:SLFC} uses the same column-wise updating scheme as  the fundamental column-wise algorithm 
shown in Section~\ref{sec:colsolve},
so that the updates to the solution vector $x$  need to be put into a critical region as well.
Finally, if we take a closer look at the execution order of the computations in this algorithm, 
we  find that this algorithm actually uses an even more ``aggressive'' scheduling scheme for  concurrency 
than the row-wise self-scheduling algorithm, which is illustrated in the example shown in Figure~\ref{fig:triag}.

\begin{algorithm}[htbp]
\caption {SLFC: Column-wise forward substitutions with self-scheduling\label{alg:SLFC}}
\begin{algorithmic}[1]
\FOR[columns in parallel]{$i=1,2,\ldots,n$}
\WHILE{$count(i) \neq 0$ }
\STATE \COMMENT{do nothing}
\ENDWHILE
\STATE $x(i) := x(i) / d(i)$
\FOR{$j=ib(i),\ldots,ib(i+1)-1$}
\STATE CRITICAL SECTION ENTRY
\STATE $x(jb(j)) = x(jb(j)) - b(j) \times x(i)$
\STATE $count(jb(j)) := count(jb(j)) - 1$
\STATE CRITICAL SECTION EXIT
\ENDFOR
\ENDFOR
\end{algorithmic}
\end{algorithm}

In Figure~\ref{fig:triag}, we show a situation where solving  $x(i)$ depends on the solutions of $x(j_1)$, $x(j_2)$ and $x(j_3)$. 
In Algorithm~\ref{alg:SLFR}, the reduction operations (lines 5--7) for computing $x(i)$ are not started until the solutions for $j_1$, $j_2$ and $j_3$ are all available.
A more aggressive strategy is to immediately start the computations when individual 
solutions become available  and save the
partial results in $x(i)$. 
Specifically, suppose that $x(j_1)$ is
available earlier than $x(j_2)$ and $x(j_3)$. Then, the partial result 
$L_{i,j_1}  x(j_1)$ can be computed  and subtracted from $x(i)$  while waiting for the solutions of $x(j_2)$ and $x(j_3)$. 
This is essentially how the computations are scheduled in the column-wise  algorithm.
Clearly, there is a finer-level concurrency in Algorithm~\ref{alg:SLFC} 
than Algorithm~\ref{alg:SLFR}
at the price of having a bigger critical region.

\subsection{CUDA implementations}
We implemented the four aforementioned SpTrSv algorithms in CUDA for
NVIDIA GPUs, namely they are the row-wise level-scheduling algorithm
(LEVR),
the column-wise level-scheduling algorithm (LEVC), 
the row-wise self-scheduling algorithm (SLFR), 
and the column-wise self-scheduling algorithm (SLFC).
In this section, we  discuss the implementations of these CUDA kernels in turn. 
Only the implementations of forward substitution will be presented. The implementations of
backward substitution are straightforward.

We shall start with kernel \texttt{LEVR} that is a CUDA implementation of
Algorithm~\ref{alg:LEVR}. The input \texttt{x} is the solution vector that is
assumed to have been initialized as the right-hand side before the entry of this function.
Arrays
\texttt{ia}, \texttt{ja} and \texttt{a} contain the strictly lower 
triangular matrix in the CSR format, while the diagonal is stored in vector \texttt{d} separately.
Array \texttt{jlev} contains the indices of the unknowns that are
grouped by levels, and \texttt{l1} and \texttt{l2} are the pointers of the starting positions
of the current  group and the next.
The information in \texttt{jlev}, \texttt{l1} and \texttt{l2} are assumed to
have been obtained from the setup phase.

\noindent
\begin{lstlisting}
__global__ void LEVR(REAL *x, REAL *a, int *ja, int *ia, REAL *d, 
                     int *jlev, int l1, int l2) {
  int wid = (blockIdx.x*BLOCKDIM+threadIdx.x)/WARP;
  int lane = threadIdx.x & (WARP-1);
  volatile __shared__ REAL r[BLOCKDIM+16];
  if (wid >= l2-l1) return;
  int i = jlev[l1+wid];
  int p1 = ia[i], q1 = ia[i+1];
  REAL sum = 0.0;
  for (int k=p1+lane; k<q1; k+=WARP)
    sum += a[k] * x[ja[k]];
  r[threadIdx.x] = sum;  // parallel reduction
  r[threadIdx.x] = sum = sum + r[threadIdx.x+16];
  r[threadIdx.x] = sum = sum + r[threadIdx.x+8];
  r[threadIdx.x] = sum = sum + r[threadIdx.x+4];
  r[threadIdx.x] = sum = sum + r[threadIdx.x+2];
  r[threadIdx.x] = sum = sum + r[threadIdx.x+1];
  if (lane == 0)
    x[i] = (x[i] - r[threadIdx.x]) / d[i];
}
\end{lstlisting}

In the function \texttt{LEVR}, one warp of threads, a CUDA concept that means $32$ consecutive threads (i.e., $\mathtt{WARP}\equiv 32$), 
are 
dispatched for one unknown of the current level and thus access one row of the matrix
associated with their global warp index (line 7), namely \texttt{wid}.
The CUDA built-in variables \texttt{blockIdx} and \texttt{threadIdx} contain the block index and the thread index within the block, and
the constant \texttt{BLOCKDIM} is the dimension of the block.
The dot-product operations between the sparse row \texttt{i}  and vector \texttt{x} are
performed by the parallel reduction within the warp (lines 10--17) in the
array \texttt{r} located in the shared memory.
Finally, the first thread of the warp, which is the one with \texttt{lane == 0}, saves the result back to \texttt{x(i)}.

Next, we will see the CUDA implementation of Algorithm~\ref{alg:LEVC} that is shown
in the kernel function \texttt{LEVC}, where the inputs \texttt{ib}, \texttt{jb} and
\texttt{b} are the sparse lower triangular matrix in the CSC format. 
The variable \texttt{wlane} is the local warp index in the thread block.
The critical
section in Algorithm~\ref{alg:LEVC} is implemented by using the CUDA \texttt{atomicAdd} operation.
An optimization  used here on  memory transactions
is to let only the first thread of a warp compute \texttt{x(i)} and save
the result into \texttt{s\_xi} in the shared memory, and then let all the other threads of the
same warp read this value from \texttt{s\_xi}. A small performance gain has been observed by this
optimization compared with letting all the threads in the warp compute \texttt{xi}.

\noindent
\begin{lstlisting}
__global__ void LEVC(REAL *x, REAL *b, int *jb, int *ib, REAL *d, 
                     int *jlev, int l1, int l2) {
  int wid = (blockIdx.x*BLOCKDIM+threadIdx.x)/WARP;
  int lane = threadIdx.x & (WARP-1);
  int wlane = threadIdx.x / WARP;
  volatile __shared__ REAL s_xi[BLOCKDIM/WARP];
  if (wid >= l2-l1) return;
  int i = jlev[l1+wid];
  if (lane == 0) s_xi[wlane] = x[i]/d[i];
  REAL xi = s_xi[wlane];
  int p1 = ib[i], q1 = ib[i+1];
  for (int j=p1+lane; j<q1; j+=WARP)
    atomicAdd(&x[jb[j]], -xi*b[j]);
  if (lane == 0) x[i] = xi;
}
\end{lstlisting}

The above two kernel functions solve all the unknowns in one level at a time, 
so that the entire forward substitutions  can be done by repeatedly running the  functions in an outer-loop 
as shown in the following, 
since the easiest way to achieve global synchronization in CUDA is to launch a new kernel after the current one finishes. 

\noindent
\begin{lstlisting}
for (int i=0; i<nlev; i++) { /* ilev: level pointers */
  LEVR<<<...,...>>>(x, a, ja, ia, d, jlev, ilev(i), ilev(i+1); 
}
\end{lstlisting}

In the rest of this section, we  discuss the CUDA kernels that implement
the self-scheduling algorithms. The kernel function \texttt{SLFR} implements 
the row-wise self-scheduling approach in Algorithm~\ref{alg:SLFR}.
Since this algorithm requires column access to the nonzero pattern of the matrix, 
column pointers \texttt{ib} and row indices \texttt{jb}
are also provided as inputs in addition to the CSR format matrix \texttt{ia}, \texttt{ja} and \texttt{a}. 
On the entry of this function, array \texttt{dp}  contains
the count of dependencies for each unknown and it contains all zeros on exit.
The input array \texttt{jlev} has the same meaning as the one in the kernel functions \texttt{LEVR}
and \texttt{LEVC}. 

\noindent
\begin{lstlisting}
__global__ void SLFR(int n, REAL *x, REAL *a, int *ja, int *ia, 
                     REAL *d, int *jb, int *ib, int *dp, int *jlev) {
  int wid = (blockIdx.x*BLOCKDIM+threadIdx.x)/WARP;
  int lane = threadIdx.x & (WARP-1);
  volatile __shared__ REAL r[BLOCKDIM+16];
  REAL ti, xi, sum = 0.0;
  if (wid >= n) return;
  int i = jlev[wid];
  int p1 = ia[i], q1 = ia[i+1];
  int p2 = ib[i], q2 = ib[i+1];
  if (lane == 0) {
    ti = 1.0 / d[i];  xi = x[i];
    while (((volatile int *)dp)[i] != 0);
  }
  for (int k=p1+lane; k<q1; k+=WARP)
    sum += a[k]*x[ja[k]];
  r[threadIdx.x] = sum;   // parallel reduction
  r[threadIdx.x] = sum = sum + r[threadIdx.x+16];
  r[threadIdx.x] = sum = sum + r[threadIdx.x+8];
  r[threadIdx.x] = sum = sum + r[threadIdx.x+4];
  r[threadIdx.x] = sum = sum + r[threadIdx.x+2];
  r[threadIdx.x] = sum = sum + r[threadIdx.x+1];
  if (lane == 0) {
    x[i] = (xi - r[threadIdx.x]) * ti;
    __threadfence();        
  }
  for (int j=p2+lane; j<q2; j+=WARP)
    atomicSub(&dp[jb[j]], 1);
}
\end{lstlisting}

In the function \texttt{SLFR}, the busy-waiting loop is at line 13, where the counter
of unknown $x(i)$ is constantly being compared with zero. 
The keyword \texttt{volatile} is used to tell the compiler that this variable can be changed at any time by other threads \cite{CUDAGuide}. 
Once the warp breaks out of the busy loop, it proceeds with the parallel reduction  as in \texttt{LEVR}. 
In line 28, the counters of the unknowns that depend on
\texttt{i} are decreased by one using the \texttt{atomicSub} operation.
Note that a ``thread fence'' is put up after 
updating  \texttt{x[i]} by the first thread of a warp and before decreasing  the counters.
This is to guarantee that at the time when the threads working on the unknowns that depend on \texttt{i}
see the corresponding counters equal to zeros, the updates to \texttt{x[i]} have already been observed by all the threads in the device \cite{CUDAGuide}.

Finally, we  discuss the implementation of the column-wise self-scheduling approach 
in Algorithm~\ref{alg:SLFC}, which is given in 
the kernel function \texttt{SLFC}. 
The CSC format matrix in \texttt{ib}, \texttt{jb} and \texttt{b} is given 
as an input and the other inputs and outputs are the same as those of \texttt{SLFR}. 
The function \texttt{SLFC} uses the same counter-based busy-waiting 
scheme as in \texttt{SLFR} and uses the same implementation of the column-wise updates to $x$ 
as in \texttt{LEVC}. Again, as in \texttt{LEVC}, the shared memory array  \texttt{s\_xi}  is used to save the result of \texttt{xi}  computed by
the first thread of a warp.

\noindent
\begin{lstlisting}
__global__ void SLFC(int n, REAL *x, REAL *b, int *jb, int *ib, 
                     REAL *d, int *dp, int *jlev) {
  int wid = (blockIdx.x*BLOCKDIM+threadIdx.x)/WARP;
  int wlane = threadIdx.x / WARP;
  int lane = threadIdx.x & (WARP - 1);
  volatile __shared__ REAL s_xi[BLOCKDIM/WARP];
  if (wid >= n) return;
  int i = jlev[wid];
  int p1 = ib[i], q1 = ib[i+1];
  REAL ti, xi;
  if (lane == 0) {
    ti = 1.0 / d[i];
    while (((volatile int *)dp)[i]);
    xi = x[i] * ti;  s_xi[wlane] = xi;
  }
  xi = s_dx[wlane];
  for (int j=p1+lane; j<q1; j+=WARP) {   
    atomicAdd(&x[jb[j]], -xi*b[j]);
    __threadfence();
    atomicSub(&dp[jb[j]], 1);
  }
  if (lane == 0) x[i] = xi;
}
\end{lstlisting}

Before closing this section, we remark that
a similar method to Algorithm~\ref{alg:SLFC} was  adopted in the ``global-synchronization-free'' algorithm 
proposed  in
\cite{Liu2016}. We list the major differences between the two algorithms and their implementations as follows:

\begin{enumerate}
\item In the forward substitution in \cite{Liu2016}, a warp is dispatched
to the unknown with the global index of the warp, i.e., \texttt{wid} in functions  \texttt{SLFR} and \texttt{SLFC}.
However, in our algorithms, the mapping between warps and the unknowns respects 
the order in \texttt{jlev}, where the unknowns are listed in a non-descending order of their levels 
(see line 8 of \texttt{SLFR} and \texttt{SLFC} for the assignments of the warps to the unknowns).
From our experimental results, we found that this warp-unknown mapping  can often significantly reduce the waiting time and thus can yield much higher overall performance.
On the negative side, our algorithms require the level information as input which is not needed by the
algorithms in \cite{Liu2016}. Consequently, the setup phase of our algorithms will be more costly for computing the levels.

\item In our algorithms, since warps with consecutive indices in a CUDA block in general do not work on consecutive unknowns, 
the optimization used in \cite{Liu2016} that utilizes the shared memory to perform partial updates to
the dependent counters in \texttt{dp} and the solution \texttt{x} cannot be applied directly. However, this
optimization can still be used if we preorder the matrix
columns according to the levels of the corresponding unknowns.
Nevertheless, in this work we do not assume that the original matrix has been reordered.

\item In our implementations of the self-scheduling algorithms, only the first thread of a
warp is spinning on the lock (line 13 in \texttt{SLFR} and \texttt{SLFC}), whereas the other
threads of the warp will wait for the first thread 
to break its busy-waiting loop before proceeding to the later instructions.
This coordination is actually guaranteed by the warp-level synchronization, that is, a warp can only execute one common instruction at a time and warp divergence is serialized.
On the other hand, in \cite{Liu2016}, the whole warp is
spinning.
\end{enumerate}

It turns out that the above differences have significant performance effects.
As we show in the experimental results in Section~\ref{sec:num}, our implementations of the self-scheduling algorithms  
exhibit superior performance compared with the implementation of the synchronization-free algorithm in \cite{Liu2016}.

\subsection{Setup phases} \label{sec:analy}
All the parallel SpTrSv algorithms discussed in this paper require a setup phase, 
where the parallelism in the solve phase is discovered from the nonzero pattern of the sparse triangular matrix.
The justification of paying the extra cost at the setup phase but having a more efficient solve phase has been discussed 
at the end of Section~\ref{sec:intro}.
In this section, we  examine the operations required in the setup phases of the algorithms discussed in the previous section.
Table~\ref{tab:setup} tabulates the major steps in the setup phases, where the first column lists the setup phases of the different
SpTrSv algorithms. ``*\_CPU'' and ``*\_GPU'' indicates  whether the setup phase is running on the CPU or on the GPU.

\begin{table}[ht]
\caption{The operations required in the setup phases of different SpTrSv algorithms \label{tab:setup}}
\begin{center}
{\small
\renewcommand{\arraystretch}{1.19}
\tabcolsep1mm
\vspace{-1em}
\begin{tabular}{r|cccccc}
 & MAT\_D2H & TRANS\_GPU & DEP\_GPU & LEV\_CPU & LEV\_GPU & LEV\_H2D\tabularnewline
 \hline
LEVR\_CPU & \checkmark &  &  & \checkmark &  & \checkmark \tabularnewline
LEVR\_GPU &  & \checkmark & \checkmark &  & \checkmark & \tabularnewline
\hline 
LEVC\_CPU & \checkmark &  &  & \checkmark &  & \checkmark \tabularnewline
LEVC\_GPU &  &  & \checkmark &  & \checkmark & \tabularnewline
\hline 
SLFR\_CPU & \checkmark & \checkmark & \checkmark & \checkmark &  & \checkmark \tabularnewline
SLFR\_GPU &  & \checkmark & \checkmark &  & \checkmark & \tabularnewline
\hline 
SLFC\_CPU & \checkmark &  & \checkmark & \checkmark &  & \checkmark \tabularnewline
SLFC\_GPU &  &  & \checkmark &  & \checkmark & \tabularnewline
\hline 
GSF\_GPU &  &  & \checkmark &  &  & \tabularnewline
\hline 
\end{tabular}
}
\end{center}
\end{table}

In the setup phases of the level-scheduling algorithms, namely \texttt{LEVR} and \texttt{LEVC}, the major computation is to decide the level
of each unknown. To do this computation on the CPU, indicted by ``LEV\_CPU'' in the table, the implementations are straightforward. They are simply 
the two for-loops
presented at the beginning of Section~\ref{sec:levlsched}, for matrices in row-wise storage formats and column-wise storage formats respectively. 
These two for-loops require operations of the order of the number of the nonzeros of the matrix and are generally hard to parallelize.
Moreover, when the level information is computed on the host, memory transfers between the host and the device are also needed, since the matrices are assumed to 
be stored  on the device and the computed array $jlev$ is required on the device.
The memory copies are denoted by ``MAT\_D2H'' and ``LEV\_H2D'' in the table.

In the setup phases of the self-scheduling algorithms,  \texttt{SLFR} and \texttt{SLFC}, 
computing the level information is also involved.
Additionally, the numbers of the dependents of all the unknowns are required as well, 
which can be easily obtained by counting the number of nonzeros per row.
This operation is denoted by ``DEP\_GPU''. Since both of the self-scheduling algorithms require column-wise access to the matrices, the setup phase
of algorithm \texttt{SLFR} also includes a transposition step, in which we transpose the CSR matrices on the GPU (nonzero pattern only without numerical values)
to obtain the matrices in the CSC format. This transposition step is denoted by ``TRANS\_GPU'' in the table.
On the other hand, the setup phase of the synchronization-free algorithm in \cite{Liu2016} is the simplest, which only
requires the dependent counters and the cost of which is almost free.

Among all these aforementioned operations in the setup phases, calculating the levels turns out to be 
the most expensive computation
compared with the costs of the other operations, which are basically negligible.
Attempts have been made to accelerate this computation by performing
a parallel topological sorting algorithm, namely Kahn's algorithm  \cite{Kahn:1962:TSL:368996.369025}, 
on GPUs, which is labeled as ``LEV\_GPU'' in Table~\ref{tab:setup}.
In the next section, we will discuss the CUDA implementation of this algorithm.

\subsubsection{Implementing Kahn's algorithm in CUDA}
The idea of Kahn's algorithm is to perform topological sorting on the DAG by repeatedly finding vertices of in-degree zero, which are called roots, saving
the roots of the current level into a queue, and 
removing the roots and their outgoing edges from the graph.
This algorithm was first described by Kahn in 1962 and also can be found in \cite{Cormen:2001}.
To the best of our knowledge, the first CUDA implementation of this algorithm was due to Naumov \cite{naumov2011parallel}
with a modified form of parallel breadth first search (BFS).

To start Kahn's algorithm, the output queue should be first initialized with the roots of level 0. 
The kernel function \texttt{FIND\_LEVEL0} shown below performs this initialization,
where \texttt{dp} contains the dependent counters, \texttt{jlev} implements the queue, and
\texttt{last} points to  the past-the-end position of the elements in the queue,
 which should be set to zero on entry.
As shown in this function, the level-0 roots are found in parallel and saved into \texttt{jlev}, and \texttt{last}
is raised by the \texttt{atomicAdd} operation.

\noindent
\begin{lstlisting}
__global__ void FIND_LEVEL0(int n, int *dp, int *jlev, int *last) {
  int gid = blockIdx.x * BLOCKDIM + threadIdx.x;
  if (gid >= n) return;
  if (dp[gid] == 0)
    jlev[atomicAdd(last, 1)] = gid;
}
\end{lstlisting}

After the initialization step for level 0, roots of the succeeding levels can be computed by the kernel
function \texttt{FIND\_LEVEL}. The input \texttt{first} is the starting position of the current level in \texttt{jlev}. 
This function requires the inputs of column pointers in \texttt{ib} 
and row indices in \texttt{jb} as well for retrieving the information of
the outgoing edges of vertices.
Therefore, the transposition is also required in the setup phase of algorithm \texttt{LEVR}, if it is running on  GPUs.
When the outgoing edges are removed, the counters in \texttt{dp} will be lowered accordingly. 
Whenever a counter reaches zero, the corresponding vertex is added to the queue and the ending position of
the current level, stored in \texttt{last}, is increased by one.

\noindent
\begin{lstlisting}
__global__ void FIND_LEVEL(int *ib, int *jb, int *dp, int *jlev, 
                           int first, int *last) {
  int wid = (blockIdx.x * BLOCKDIM + threadIdx.x) / WARP;
  int lane = threadIdx.x & (WARP - 1);
  int i = jlev[first+wid];
  int p1 = ib[i], q1 = ib[i+1];
  for (int j=p1+lane; j<q1; j+=WARP)
    if (atomicSub(&dp[jb[j]], 1) == 1)
      jlev[atomicAdd(last, 1)] = jb[j];
} 
\end{lstlisting}

Clearly, the entire topological sort can be performed by repeatedly 
calling function \texttt{FIND\_LEVEL}  until all the vertices have been removed from the graph, 
as shown in the code segment below.
The  graph is traversed one level at a time, and between two levels 
there is a synchronization point by  restarting the kernel function and 
a memory transfer from the device to the host 
for the number of the vertices that have been removed so far.
Consequently, like the level-scheduling algorithms in the solve phases,
the efficiency will gradually deteriorate with the increase of the number of  levels.
A more efficient implementation of Kahn's algorithm in CUDA without global synchronization 
 is currently being investigated by the author.

\noindent
\begin{lstlisting}
  FIND_LEVEL0<<<..., ...>>>(n, dp, jlev, last);
  cudaMemcpy(&h_last, last, sizeof(int), cudaMemcpyDeviceToHost);
  for (first=0, nlev=0, ilev[0]=0; h_last < n; ) {
    FIND_LEVEL<<<..., ...>>>(ib, jb, dp, jlev, first, last);
    cudaMemcpy(&h_last, last, sizeof(int), cudaMemcpyDeviceToHost);
    first = h_last;  ilev[nlev++] = first;
  }
  ilev[nlev] = n;
\end{lstlisting}

\section{Numerical experiments} \label{sec:num}
The experiments were conducted on one node of Ray, a Linux cluster at Lawrence Livermore National Laboratory, equipped with a
dual socket Power 8+ CPU (10 cores/socket) and 4 NVIDIA Tesla P100 (Pascal) GPUs.
The CUDA program was compiled by the NVIDIA CUDA compiler \texttt{nvcc} from CUDA toolkit 8.0 with option 
\texttt{-gencode arch=compute\_60,"code=sm\_60"} 
for compute capability 6.0,
and the CPU code was compiled by IBM C++ compiler \texttt{xlc++} with the \texttt{-O2} optimization level.
For the CUDA kernel configurations, thread blocks are one-dimensional and of size 512.
We  compared our four SpTrSv solvers that are LEVR and LEVC using the level-scheduling algorithm and SLFR and SLFC using the self-scheduling
algorithm, with the two solvers \texttt{cusparse?csrsv} and \texttt{cusparse?csrsv2} available from cuSPARSE v8.0. 
The \texttt{cusparse?csrsv} solver adopted a similar row-wise level-scheduling approach as in LEVR \cite{naumov2011parallel}, whereas
the algorithm used in \texttt{cusparse?csrsv2} has not been published.
We also considered the global-synchronization-free algorithm introduced in \cite{Liu2016}, denoted by GSF, the implementation of which is available from \cite{GSFcode}.
In the following of this section, we will first show the performance of these solvers on structured matrices obtained from finite difference
discretizations of 2-D and 3-D Laplace operators, and then we will report their performance on general  matrices.

For each test matrix $A \in \RR^{n\times n}$, letting
\begin{equation}
 A = L + D + U,
\end{equation}
where $L$ and $U$ denote the strict lower and upper triangular parts of $A$ respectively, 
and $D$ denotes the diagonal matrix that has the same
diagonal as $A$, for which we assume that $D_{ii} \neq 0$ for $i=1,\ldots,n$,
we solve the lower triangular system followed by the upper triangular system of the form
\begin{equation} \label{eq:lusolve}
y = (U+D)\inv (L+D)\inv x. 
\end{equation}
The performance of the solves in  \eqref{eq:lusolve} was measured in GFLOPS given by
\begin{equation} \label{eq:flops}
\mbox{GFLOPS} = \frac{2\times nnz}{t \times 10^9},
\end{equation}
where we denote by $nnz$ the number of nonzeros of $A$ and by $t$ the solve time  measured in seconds, which was taken as an average of 100 runs. 
Note that in \eqref{eq:flops} the factor $2$ is due to the fact that each off-diagonal nonzero entry requires
one addition and one multiplication and there are two divisions for each diagonal element.
 
For the row-wise algorithms, we assume that the matrix $A$ is stored in the CSR format on the GPU, whereas for the CSC format
we assume that $A$ is available in the CSC format on the GPU.

\subsection{2-D and 3-D Laplacians}
We shall start our performance study with a set of discretized Laplacians obtained from the finite-difference scheme on regular 
grids of the dimension $n_x \times n_y \times n_z$. The size of the matrix, denoted by $n$, is given by 
$n=n_x  n_y  n_z$.
In Table~\ref{tab:regulargrids}, we list the dimensions of the testing grids, where $\rho_{yx}$ and $\rho_{zx}$ are the aspect ratios of the grids defined as
\begin{equation} \label{eq:aspectratios}
\rho_{yx}=\frac{n_y}{n_x} \quad \mbox{and} \quad \rho_{zx}=\frac{n_z}{n_x} \ .
\end{equation}
In general, the numbers of the levels in the sparse triangular systems of the same size will increase with these ratios and thus the degrees of parallelism will decrease accordingly.
Therefore, lower performance will be expected for the grids with higher aspect ratios.
In this set of experiments, we will examine the performance behaviors of the 7 considered solvers across the testing grids.
For the 2-D grids, we tested Laplacian matrices with standard $5$- and $9$-point stencils, while for the 3-D grids we tested the standard $7$- and $27$-point Laplacian matrices. The number of stencil points is precisely the number of nonzeros of a matrix row that does not correspond to 
a grid point on the boundary. Moreover, this number is also the same as the number of the diagonals of the matrix under the assumption that
the lexicographical ordering is used for the grid points.

\begin{table}[ht]
\caption{2-D/3-D regular grids for generating Laplacian matrices \label{tab:regulargrids}}
\begin{center}
\renewcommand{\arraystretch}{1.05}
\begin{tabular}{c|c||c|c}
\multicolumn{1}{c|}{$n_x \times n_y$} & $\rho_{yx}$ & \multicolumn{1}{c|}{$n_x \times n_y \times n_z$} & $\rho_{zx}$\tabularnewline
\hline 
$1024\times1024$ & 1 & $128\times128\times128$ & 1\tabularnewline
  $512\times2048$  & 4 & $64\times128\times256$  & 4\tabularnewline
  $256\times4096$ & 16 & $64\times64\times512$   & 8\tabularnewline
  $128\times8192$ & 64 & $32\times64\times1024$  & 32\tabularnewline
  $64\times16384$ & 256 & $32\times32\times2048$ & 64\tabularnewline
\hline 
\end{tabular}
\end{center}
\end{table}

\begin{figure}[ht]
\caption{Performance of the solve phases of the SpTrSv algorithms for 2-D Laplacians on
regular grids in single  and double  precisions measured in GFLOPS \label{fig:lapsolve}}
 \begin{subfigure}[t]{0.49\textwidth}
 \caption{5-point stencil, single precision}
  \includegraphics[width=\textwidth]{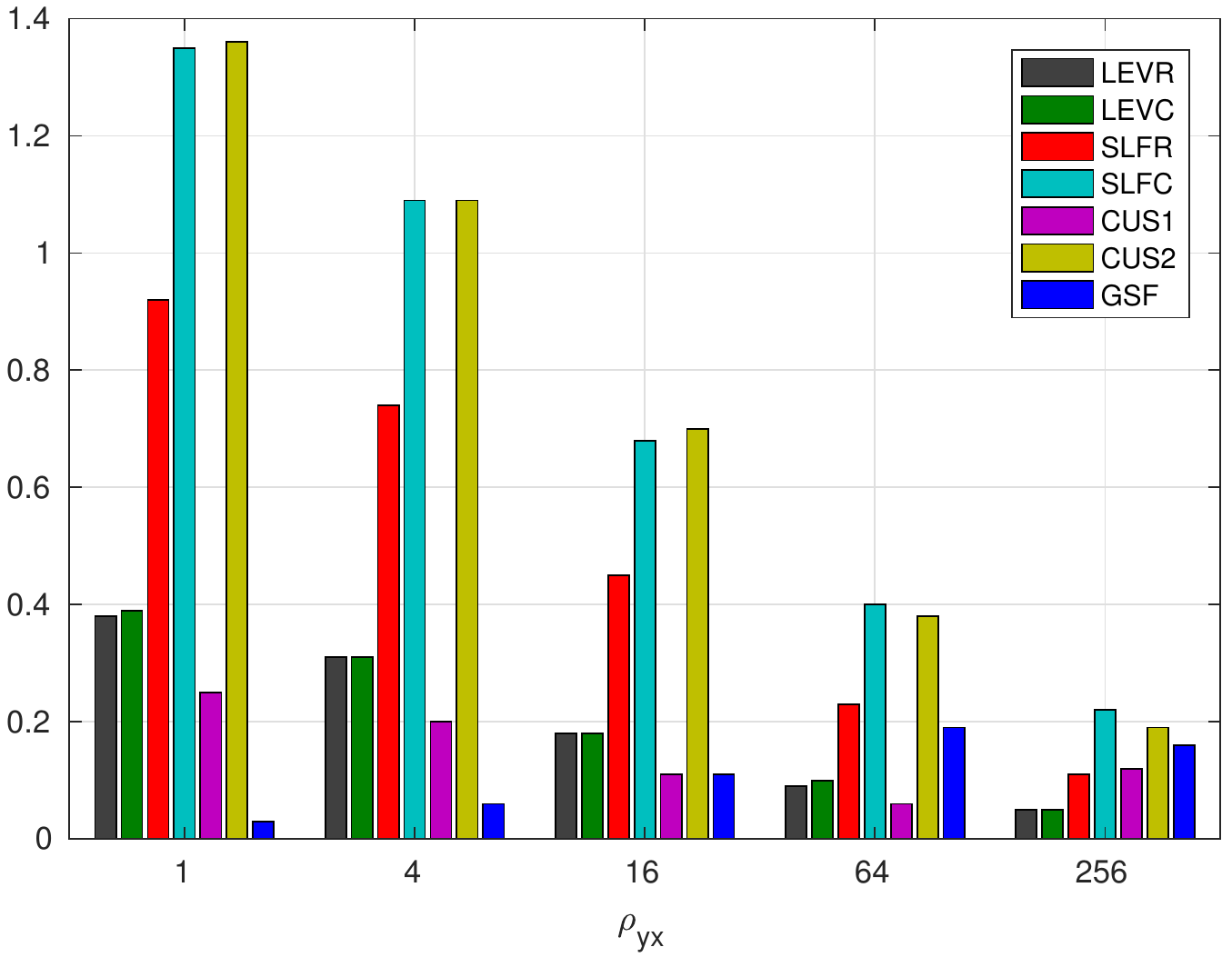}
   \end{subfigure}
 \begin{subfigure}[t]{0.49\textwidth}
 \caption{5-point stencil, double precision}
   \includegraphics[width=\textwidth]{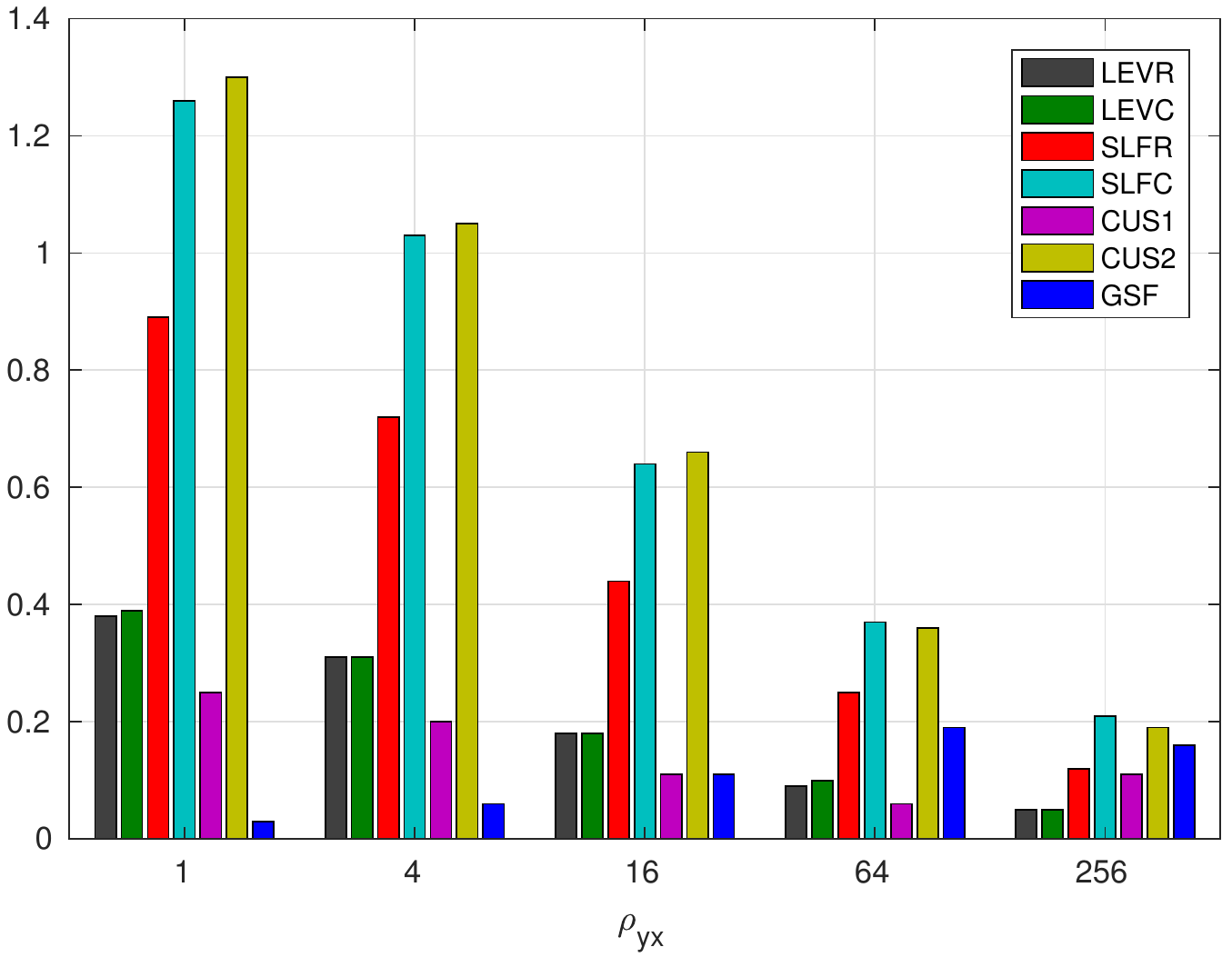}
     \end{subfigure}
     
 \begin{subfigure}[t]{0.49\textwidth}
 \caption{9-point stencil, single precision}
 \includegraphics[width=\textwidth]{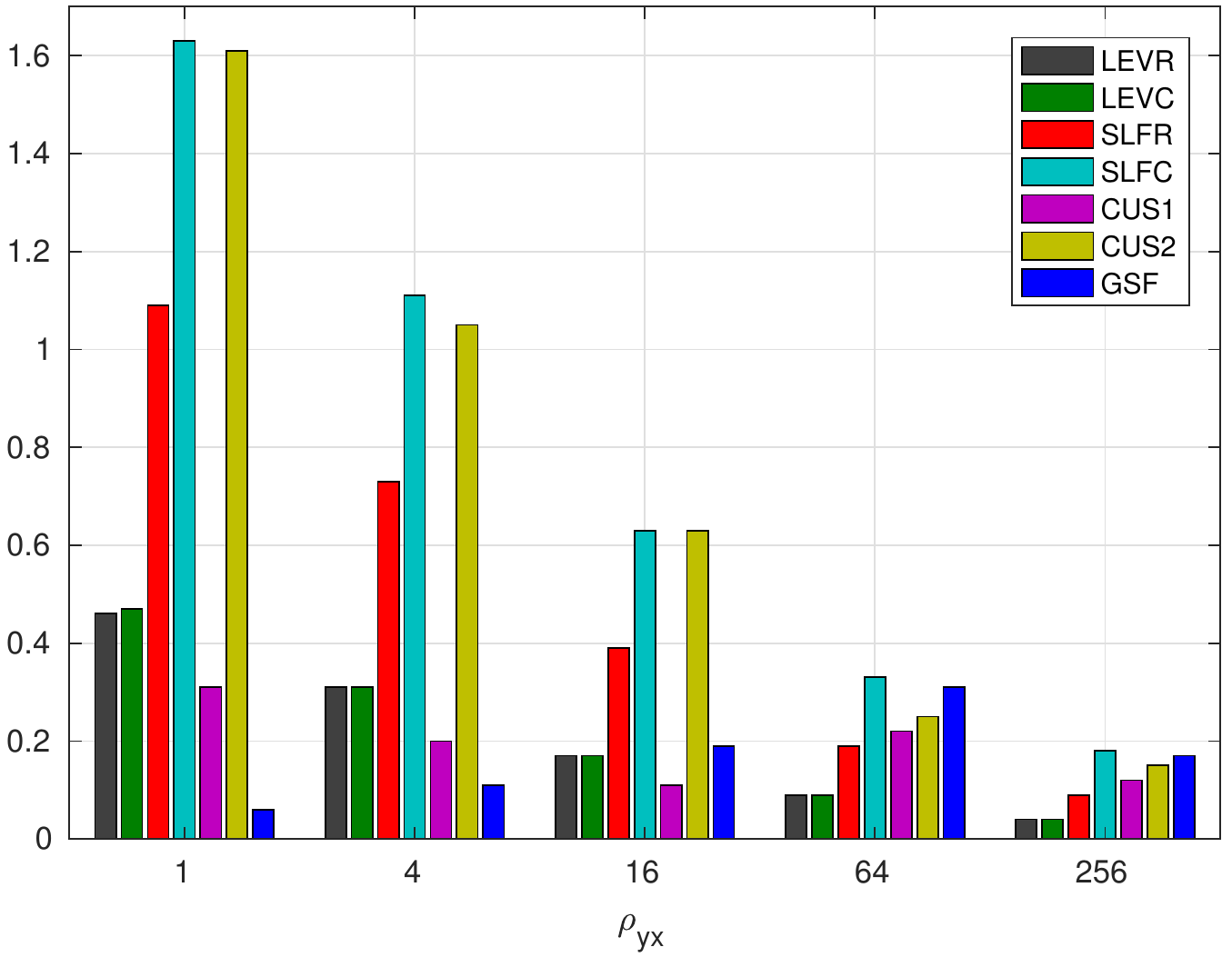}
 \end{subfigure}
 \begin{subfigure}[t]{0.49\textwidth}
 \caption{9-point stencil, double precision}
 \includegraphics[width=\textwidth]{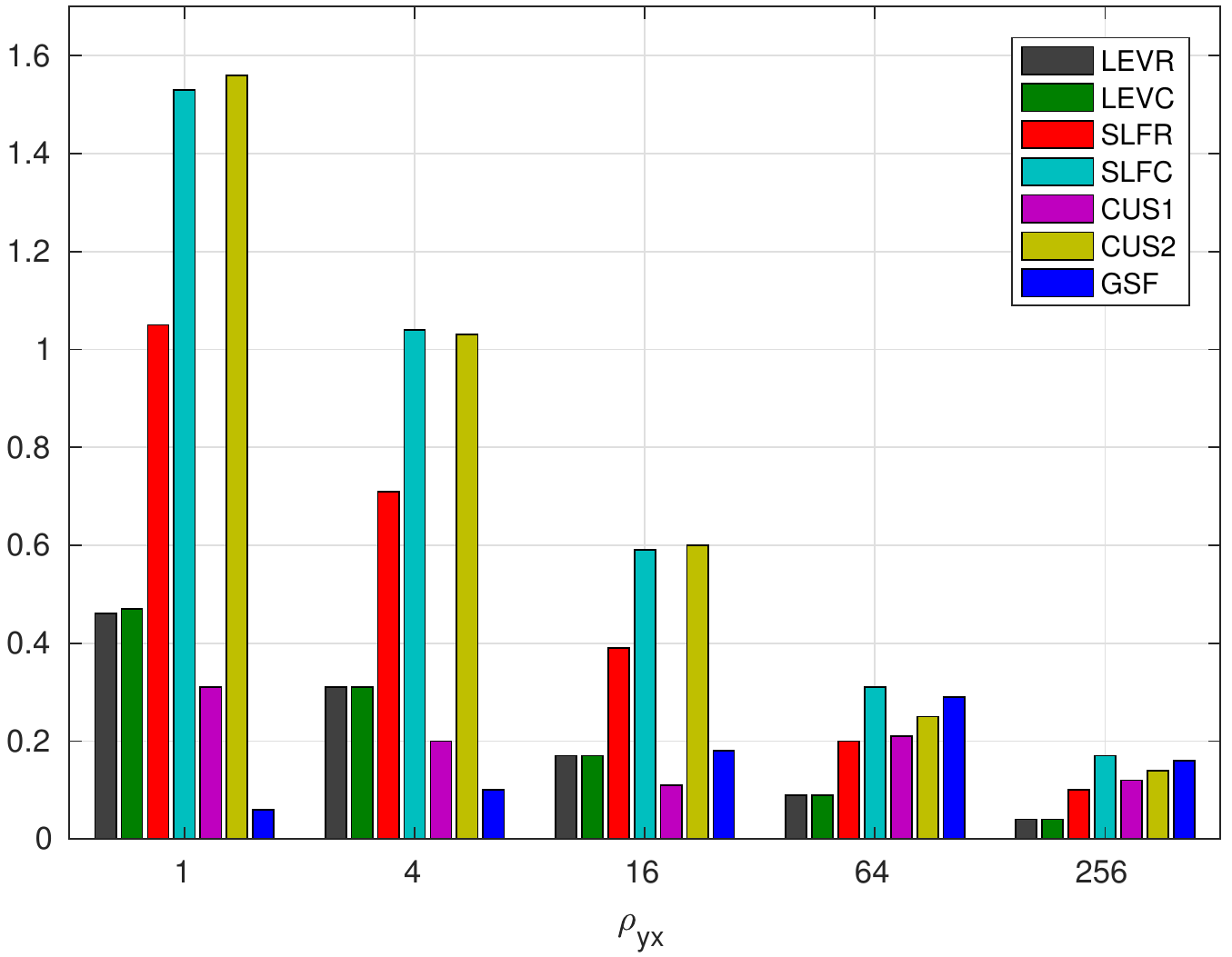}
 \end{subfigure}    
\end{figure}

\begin{figure}[ht]
\caption{Performance of the solve phases of the SpTrSv algorithms for 3-D Laplacians on
regular grids in single  and double  precisions measured in GFLOPS \label{fig:lap3dsolve}}
 \begin{subfigure}[t]{0.49\textwidth}
 \caption{7-point stencil, single precision}
  \includegraphics[width=\textwidth]{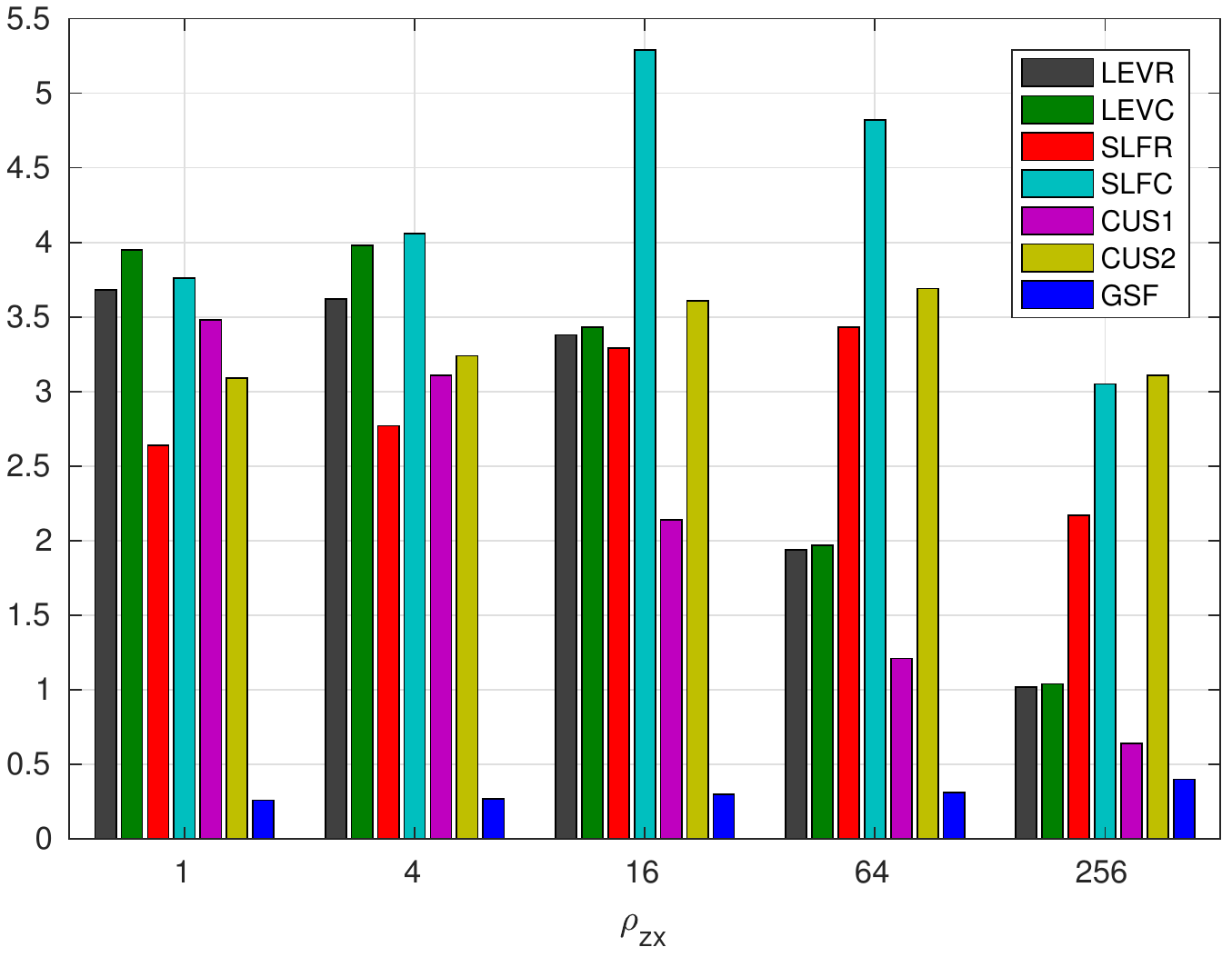}
   \end{subfigure}
 \begin{subfigure}[t]{0.49\textwidth}
 \caption{7-point stencil, double precision}
   \includegraphics[width=\textwidth]{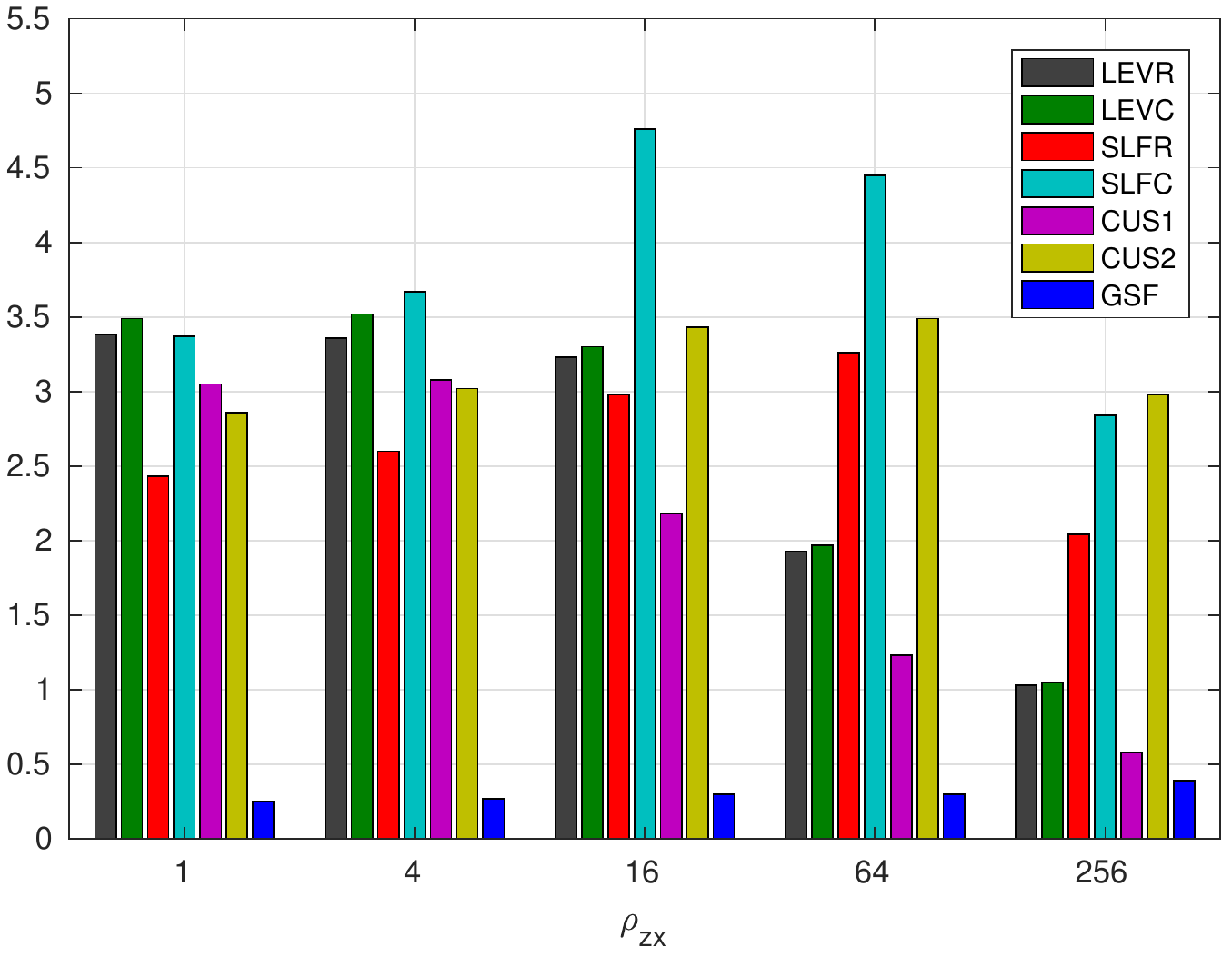}
     \end{subfigure}
     
 \begin{subfigure}[t]{0.49\textwidth}
 \caption{27-point stencil, single precision}
 \includegraphics[width=\textwidth]{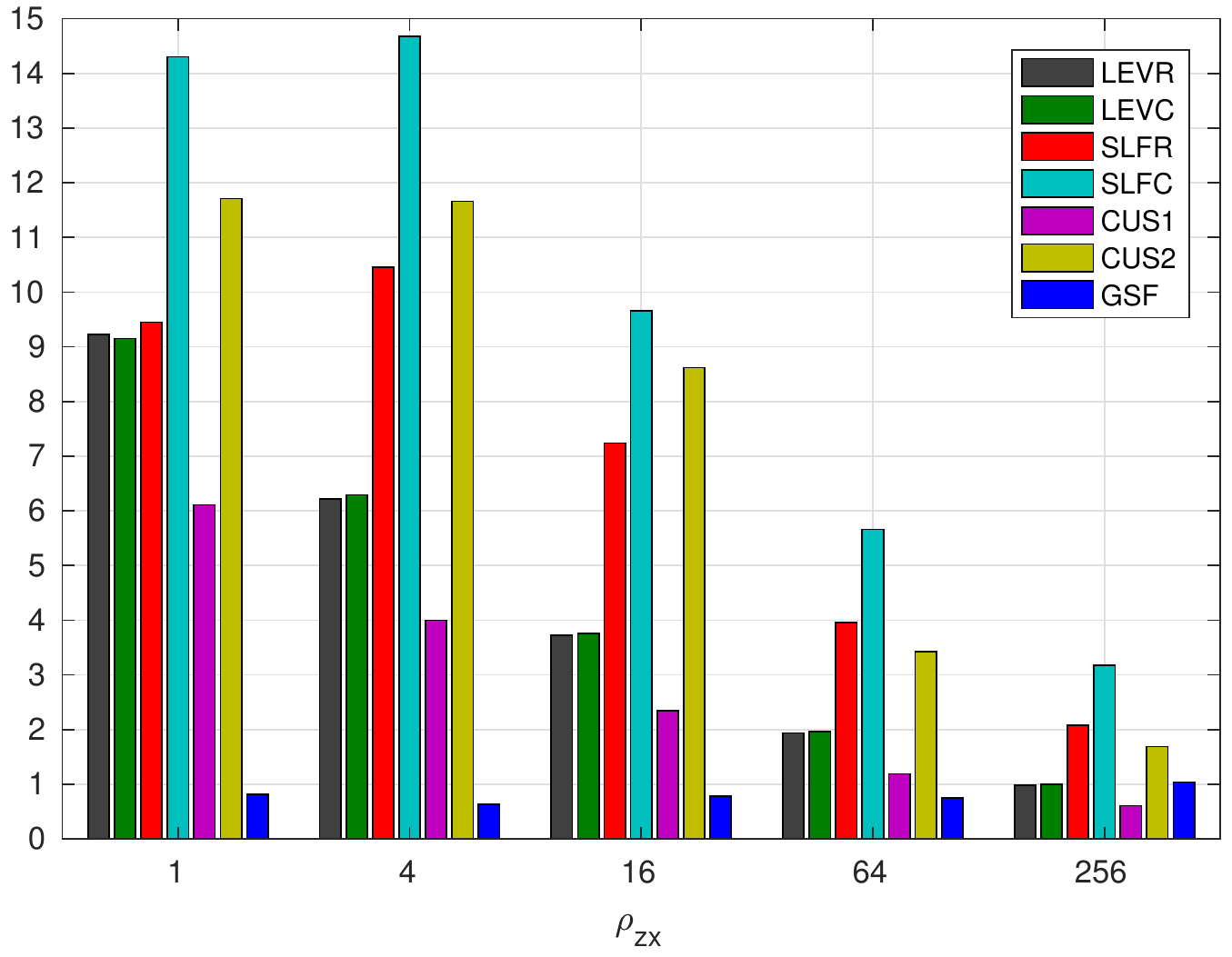}
 \end{subfigure}
 \begin{subfigure}[t]{0.49\textwidth}
 \caption{27-point stencil, double precision}
 \includegraphics[width=\textwidth]{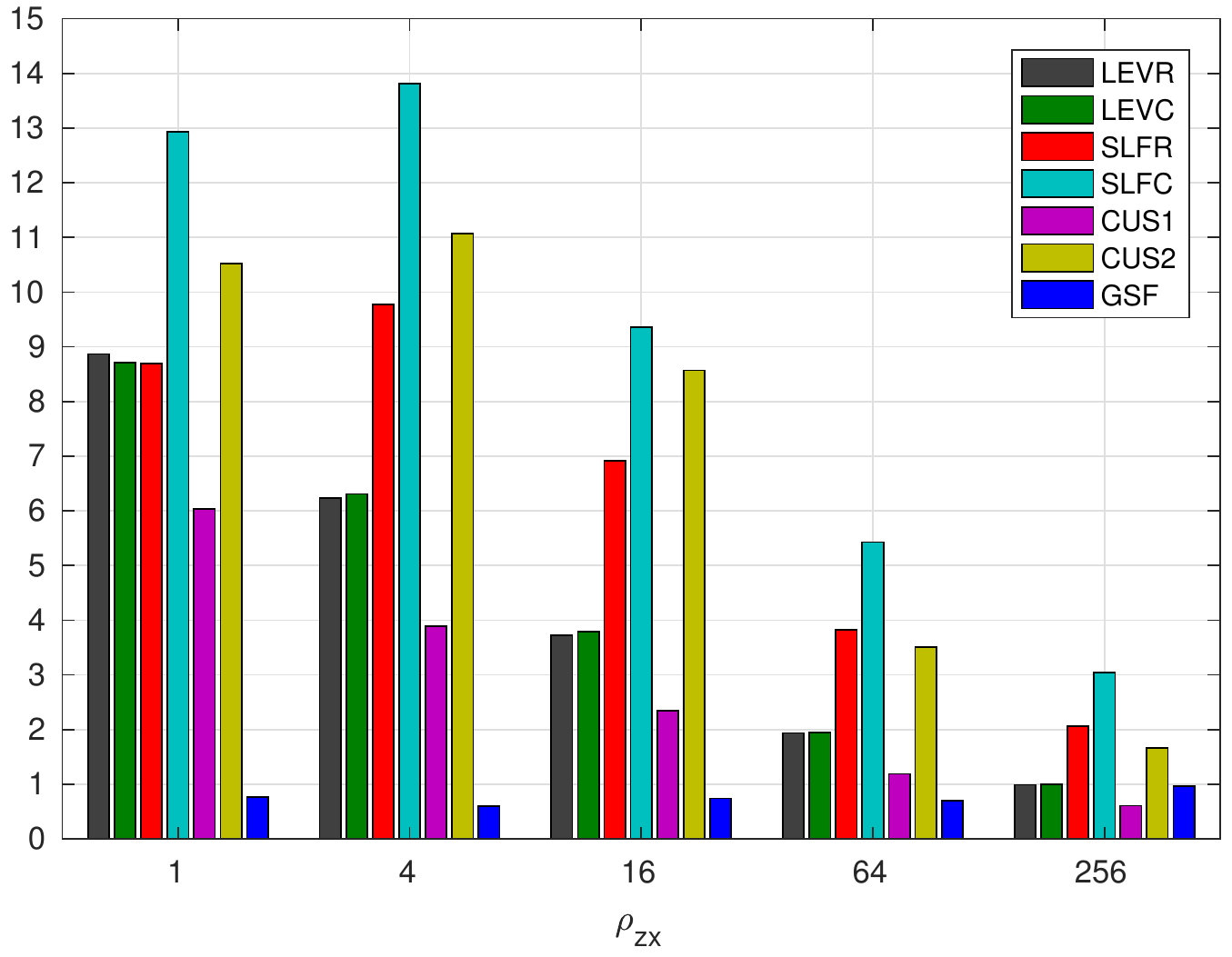}
 \end{subfigure}    
\end{figure}

In Figures \ref{fig:lapsolve} and \ref{fig:lap3dsolve}, we present the performance of the solve phases of
the triangular solves in \eqref{eq:lusolve} that is measured in GFLOPS in both single and double precisions
for  2-D and 3-D Laplacians on the regular grids listed in Table~\ref{tab:regulargrids}.
The solvers \texttt{cusparse?\_csrsv} and \texttt{cusparse?\_csrsv2} 
from cuSPARSE
are denoted by \texttt{CUS1} and \texttt{CUS2} respectively in the figures.
The x-axes of the figures represent the aspect ratios $\rho_{yx}$ and $\rho_{zx}$
that are defined in \eqref{eq:aspectratios}.
For the results of the 2-D Laplacians in Figure~\ref{fig:lapsolve}, for most of the cases, \texttt{LEVR} and \texttt{LEVC}
outperformed the cuSPARSE counterpart \texttt{cusparse?\_csrsv}, and the self-scheduling solvers \texttt{SLFR} and \texttt{SLFC} 
showed superior performance to the level-scheduling solvers. For all the grids, the column-wise algorithm \texttt{SLFC}
exhibited better performance than the row-wise algorithm \texttt{SLFR} and the performance of \texttt{SLFC} was very close to
that of \texttt{cusparse?\_csrsv2}. As expected, the performance of all the first 6 solvers steadily degraded when the ratios
$\rho_{yx}$ and $\rho_{zx}$ were increasing. On the other hand, \texttt{GSF} was the exception, which performed exceedingly poorly
for the first two grids but eventually became competitive again for the last two grids.
For all the grids and all the solvers, the performance difference between the kernels in  single precision and  double precision
was very small.

The performance for the 3-D Laplacians are shown in Figure~\ref{fig:lap3dsolve}. 
Compared with the 2-D problems, the  GFLOPS numbers
are much higher. This is because first  there are much fewer levels in the triangular matrices
for these 3-D problems,
and second there are more operations in each task due to the more complex  stencils.
When comparing the different solvers, 
the level-scheduling approaches \texttt{LEVR} and \texttt{LEVC} were faster than the row-wise self-scheduling solver \texttt{SLFR}
for most of the 7-point operators, while
the column-wise  solver \texttt{SLFC} remained 
the fastest  for all the test cases
except for the 7-point operator on the last 3-D grid, where \texttt{cusparse?\_csrsv2} slightly outperformed \texttt{SLFC}
(3.11  vs. 3.05 GFLOPS in single precision and 2.98  vs. 2.84 GFLOPS in double precision).
For the other cases, the speedup of \texttt{SLFC} over \texttt{cusparse?\_csrsv2} was 
by a factor of from $1.18$ to $1.47$ for the 7-point matrices and from $1.09$ to $1.82$ for the 27-point matrices.
For almost all the 3-D Laplacians, the performance of the \texttt{GSF} solver was lower than those of the other
solvers, and in particular the performance gap between the \texttt{GSF} solver and the self-scheduling solvers was quite significant.
Finally, we remark that the performance discrepancy between single and double precisions was more noticeable
for the 3-D problems than that for the 2-D problems.

\subsection{General matrices}
In this section, we report the performance of the SpTrSv solvers in CUDA on $10$ general matrices
selected from the University of Florida sparse matrix
collection \cite{Davis:UFM} and $2$ matrices for linear elasticity problems discretized 
with linear elements generated by
the finite element package MFEM \cite{mfem-library}. 
The size (N), the number of the nonzeros (NNZ), the average number of non-zeros per row (RNZ), 
and a short description of each matrix are tabulated in Table \ref{tab:gpumatrices}.
The sizes of these matrices range from half a million to 4 million, and the densities of the matrices 
vary from a few nonzeros per row to several scores.

\begin{table}[ht]
\caption[Test matrices.]{Names, orders (N), numbers of nonzeros (NNZ), 
 average numbers of nonzeros per row (RNZ), and short descriptions of the test matrices.\label{tab:gpumatrices}}
\begin{center}
\renewcommand{\arraystretch}{1.05}
\begin{tabular}{l|rrrl}
\multicolumn{1}{c|}{Matrix} & \multicolumn{1}{c}{N} & \multicolumn{1}{c}{NNZ} & \multicolumn{1}{c}{RNZ}  & \multicolumn{1}{c}{DESCRIPTION} \tabularnewline
\hline
af\_shell8 & 504,855  & 17,588,875  & 34.8 & Sheet metal forming simulation\tabularnewline
ecology2 & 999,999  & 4,995,991 & 5.0 & Landscape ecology problem \tabularnewline
webbase1M & 1,000,005 & 3,105,536 & 3.1 & Web connectivity matrix \tabularnewline
elasticity2D & 1,002,528 & 14,012,744 & 14.0 & 2D FEM elasticity \tabularnewline
elasticity3D & 1,029,000 &  80,990,208 & 78.7 & 3D FEM elasticity \tabularnewline
thermal2 & 1,228,045 & 8,580,313 & 7.0 & Steady state thermal problem \tabularnewline
atmosmodd & 1,270,432 & 8,814,880 & 6.9 & Atmospheric modeling problem \tabularnewline
StocF-1465 & 1,465,137 & 21,005,389 & 14.3 & Flow in porous medium problem \tabularnewline
af\_shell10 & 1,508,065 & 52,672,325 & 34.9 & Sheet metal forming simulation \tabularnewline
Transport & 1,602,111  & 23,500,731 & 14.7 & 3D FEM flow and transport \tabularnewline
Bump\_2911 & 2,911,419 &  127,729,899 & 43.8 & 3D  reservoir simulation \tabularnewline
Queen\_4147 & 4,147,110 & 329,499,284   & 79.4 & 3D structural problem \tabularnewline
\hline
\end{tabular}
\end{center}
\end{table}

\begin{figure}[ht]
\caption{The numbers of the levels (NLEV) in the lower and upper triangular parts of the tested matrices 
and the performance of the solve phases of the SpTrSv algorithms, 
measured in GFLOPS and in double precision.\label{fig:genmat} }
\centerline{\includegraphics[width=\textwidth]{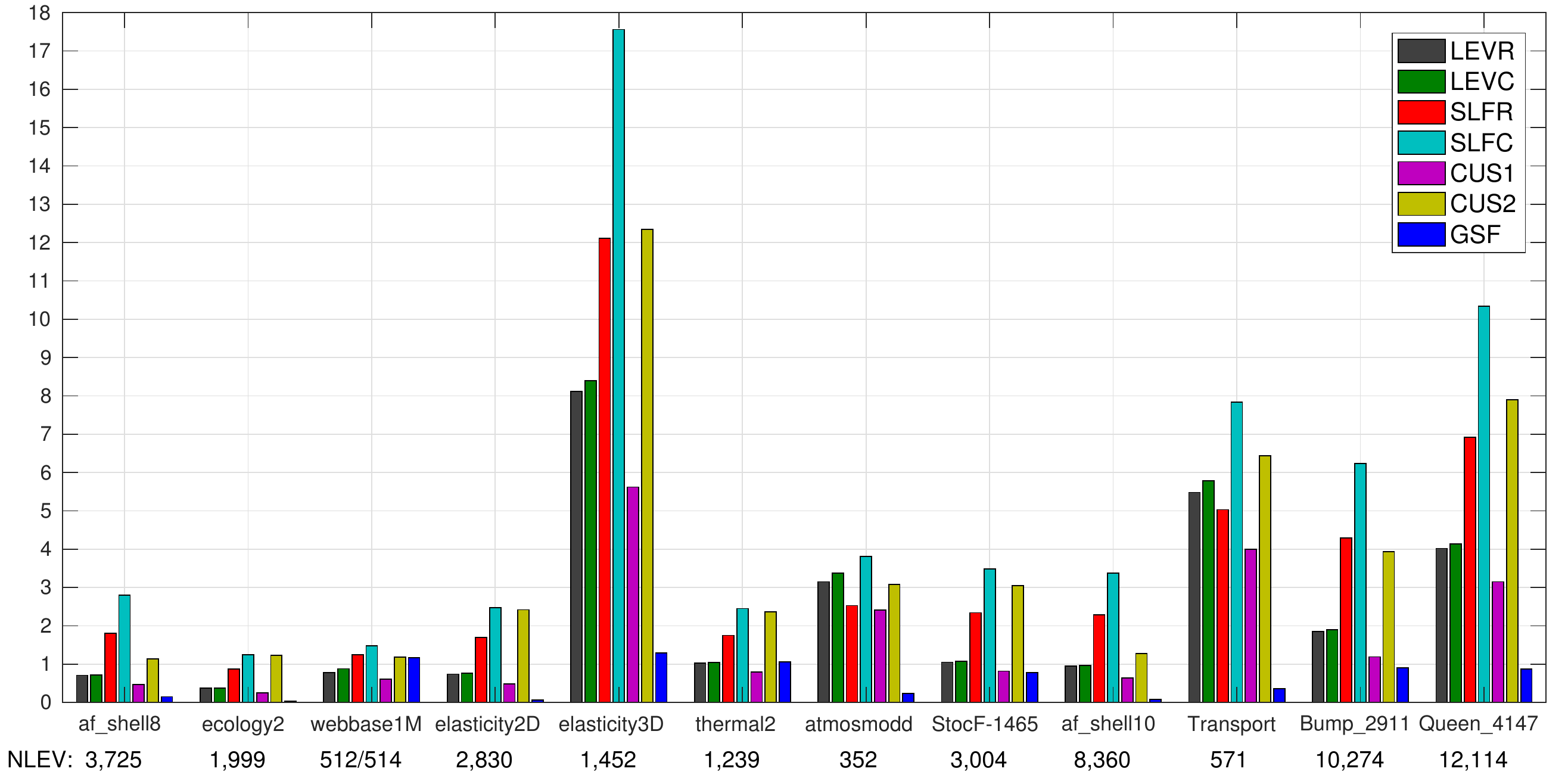}}
\end{figure}

In Figure~\ref{fig:genmat}, we report the performance of the 7 tested solvers in double precision only.
The numbers of the levels in the lower and upper triangular parts of the matrices are presented at the bottom of
this figure.
As shown, our \texttt{LEVR} and \texttt{LEVC} solvers 
consistently outperformed the cuSPARSE counterpart \texttt{cusparse?\_csrsv}, and our \texttt{SLFC} solver
achieved speedup over \texttt{cusparse?\_csrsv2} by a factor of up to $2.64$.
The highest GFLOPS numbers ($>10$ GFLOPS) were achieved with matrices \texttt{elasticity3D} and \texttt{Queen\_4147}, 
both of which have denser rows (about 79 nonzeros per row) than the other matrices.
Note that although there are a large number of levels in \texttt{Queen\_4147},
the self-scheduling algorithms still reached high performance in the solves.
Lastly,  for most of the matrices, 
the performance of the \texttt{GSF} solver was
not very competitive compared with the other solvers.

\subsection{Cost of the setup phases}

Our final experimental results to report are the timings of the setup phases of
all the sparse triangular system solvers considered in this paper.
In Table~\ref{tab:lapsetup}, we present the timings required by these
setup phases running the CPU, which were measured in milliseconds.
In this table, the numbers in the parentheses are the ratios between the setup time and the time for one solve of
\eqref{eq:lusolve} in double precision.
As shown, the setup time for the 2-D problems was generally small, not much more expensive than the solve time.
Moreover, the timings for running the setup phases on the CPU was almost constant for the various shapes of grids.
Since the performance of the solve phases gradually degrades for grids with higher aspect ratios, the cost of the setup phase
became more and more insignificant relative to the cost of the solve phase.
The time differences in the setup phases  
between \texttt{LEVR} and \texttt{LEVC} and between \texttt{SLFR} and \texttt{SLFC}
were due to the two versions of the sequential algorithms running the CPU for computing the levels of the unknowns.
Based on the experimental results, the column-wise version was often faster than the row-wise version except for the 2-D 5-point operators.
On the other hand, the setup costs for the 3-D problems were much higher due to the more complex stencils.
Compared with the cost of the setup phase of \texttt{cusparse?\_csrsv}, the setup phases of our solvers were always cheaper
for all problems when running the CPU.
The cost of the setup phase of \texttt{cusparse?\_csrsv2} was in general the most inexpensive, especially
for the 3-D problems, among all the solvers
except for \texttt{GSF}, whereas
the setup cost of \texttt{GSF} was basically negligible but 
the performance of its solve phase has been shown to be usually much lower
than those of the others.

\begin{table}[ht]
\caption{Timings of the setup phases of the SpTrSv algorithms running on the CPU 
for the 2-D and 3-D Laplacians on regular grids, measured in milliseconds.
The numbers in the parentheses are the ratios between the setup time and the time for one solve of
\eqref{eq:lusolve} in double precision.
\label{tab:lapsetup}}
\begin{subtable}[ht]{\textwidth}
\caption{2-D Laplacians with 5-point stencil }
\begin{center}
\renewcommand{\arraystretch}{1.05}
\begin{tabular}{rrrr|rrr}
\multicolumn{1}{c}{LEVR} & \multicolumn{1}{c}{LEVC} & 
\multicolumn{1}{c}{SLFR} & \multicolumn{1}{c|}{SLFC} & 
\multicolumn{1}{c}{CUS1} & \multicolumn{1}{c}{CUS2} & 
\multicolumn{1}{c}{GSF} \tabularnewline
\hline
\phantom{0}30 (1.1) & \phantom{0}41 (1.5) & \phantom{0}34 (2.9) & \phantom{0}41 (5.0) & \phantom{0}86 (2.1) & 214 (27.) & .34 (.00) \tabularnewline
30 (.87) & 41 (1.2) & 34 (2.4) & 42 (4.1) & 94 (1.8) & 75 (7.5) & .34 (.00) \tabularnewline
30 (.51) & 41 (.72) & 34 (1.5) & 41 (2.5) & 128 (1.4) & 36 (2.3) & .35 (.00)\tabularnewline
30 (.27) & 41 (.38) & 34 (.81) & 41 (1.5) & 203 (1.2) & 25 (.87) & .34 (.00) \tabularnewline
30 (.14) & 41 (.19) & 34 (.40) & 41 (.85) & 357 (3.8) & 35 (.64) & .35 (.00) \tabularnewline
\hline 
\end{tabular}
\end{center}
\end{subtable}


\begin{subtable}[t]{\textwidth}
\caption{2-D Laplacians with 9-point stencil}
\begin{center}
\renewcommand{\arraystretch}{1.05}
\begin{tabular}{rrrr|rrr}
\multicolumn{1}{c}{LEVR} & \multicolumn{1}{c}{LEVC} & 
\multicolumn{1}{c}{SLFR} & \multicolumn{1}{c|}{SLFC} & 
\multicolumn{1}{c}{CUS1} & \multicolumn{1}{c}{CUS2} & 
\multicolumn{1}{c}{GSF} \tabularnewline
\hline 
\phantom{0}45 (1.1) & \phantom{0}36 (.90) & \phantom{0}52 (2.9) & \phantom{0}36 (2.9) & 120 (2.0) & 208 (17.) & .41 (.00) \tabularnewline
45 (.71) & 36 (.59) & 52 (2.0) & 37 (2.0) & 151 (1.6) & 74 (4.0) & .40 (.00) \tabularnewline
44 (.39) & 36 (.32) & 52 (1.1) & 36 (1.1) & 223 (1.3) & 37 (1.2) & .40 (.00) \tabularnewline
44 (.20) & 36 (.17) & 52 (.54) & 36 (.60) & 377 (4.2) & 36 (.47) & .40 (.01) \tabularnewline
44 (.10) & 36 (.08) & 52 (.27) & 36 (.33) & 688 (4.4) & 79 (.61) & .39 (.00) \tabularnewline
\hline 
\end{tabular}
\end{center}
\end{subtable}


\begin{subtable}[t]{\textwidth}
\caption{3-D Laplacians with 7-point stencil}
\begin{center}
\renewcommand{\arraystretch}{1.05}
\begin{tabular}{rrrr|rrr}
\multicolumn{1}{c}{LEVR} & \multicolumn{1}{c}{LEVC} & 
\multicolumn{1}{c}{SLFR} & \multicolumn{1}{c|}{SLFC} & 
\multicolumn{1}{c}{CUS1} & \multicolumn{1}{c}{CUS2} & 
\multicolumn{1}{c}{GSF} \tabularnewline
\hline
\phantom{0}77 (8.9) & \phantom{0}65 (7.8) & \phantom{0}89 (7.4) & \phantom{0}65 (7.5) & 116 (12.) & \phantom{0}44 (4.3) & .76 (.01)\tabularnewline
75 (8.7) & 64 (7.7) & 86 (7.8) & 63 (7.9) & 121 (13.) & 55 (5.8) & .77 (.01) \tabularnewline
75 (8.4) & 64 (7.2) & 86 (8.9) & 63 (10.) & 125 (9.2) & 44 (5.2) & .76 (.01)\tabularnewline
75 (5.0) & 63 (4.3) & 86 (9.7) & 63 (9.7) & 134 (5.6) & 46 (5.6) & .75 (.01) \tabularnewline
75 (2.7) & 64 (2.3) & 86 (6.1) & 64 (6.3) & 134 (3.0) & 34 (3.5) & .75 (.01) \tabularnewline
\hline 
\end{tabular}
\end{center}
\end{subtable}


\begin{subtable}[t]{\textwidth}
\caption{3-D Laplacians with 27-point stencil}
\begin{center}
\renewcommand{\arraystretch}{1.05}
\begin{tabular}{rrrr|rrr}
\multicolumn{1}{c}{LEVR} & \multicolumn{1}{c}{LEVC} & 
\multicolumn{1}{c}{SLFR} & \multicolumn{1}{c|}{SLFC} & 
\multicolumn{1}{c}{CUS1} & \multicolumn{1}{c}{CUS2} & 
\multicolumn{1}{c}{GSF} \tabularnewline
\hline
253 (20.)  & 189 (15.)  & 285 (22.) & 189 (22.) & 289 (16.) & 64 (6.0) & 1.9 (.01) \tabularnewline
252 (14.)  & 186 (11.) & 287 (25.) & 188 (23.) & 302 (11.) & 96 (9.4) & 1.9 (.01) \tabularnewline
249 (8.3) & 187 (6.4) & 286 (17.) & 191 (16.) & 308 (6.4) & 82 (6.3) & 1.9 (.01) \tabularnewline
250 (4.3) & 189 (3.4) & 281 (9.1) & 190 (9.4) & 351 (3.8) & 108 (3.4) & 1.9 (.01) \tabularnewline
247 (2.2) & 186 (1.7) & 282 (5.2) & 189 (5.3) & 406 (2.3) & 74 (1.1)  & 1.9 (.01) \tabularnewline
\hline 
\end{tabular}
\end{center}
\end{subtable}
\end{table}

In Table~\ref{tab:lapsetupgpu2d3d}, we report the timings for the  setup phases on the GPU for the 2-D and
3-D Laplacians. Comparing with the results in Table \ref{tab:lapsetup}, we can see that the setup phases on the GPU
required more time  than the CPU versions for the 2-D cases, whereas for the 3-D problems, the costs of the setup phases 
were reduced dramatically by the GPU. A remarkable difference between the GPU setup phases and
the CPU ones is that the running time increases with the number of levels for the same reason as in the solve phases 
based on level-scheduling algorithms: the cost of (re)launching GPU kernels becomes higher when the number of  levels is larger.
The same performance behavior can be also found from the cost of the setup phase of the solver \texttt{cusparse?\_csrsv}
shown in Table~\ref{tab:lapsetup}.
On the other hand, there is no clear evidence that the observed setup cost of \texttt{cusparse?\_csrsv2} is correlated with the number of levels.

\begin{table}[ht]
\caption{Timings of the setup phases of the SpTrSv algorithms running on the GPU 
for the 2-D and 3-D Laplacians on regular grids, measured in milliseconds.
The numbers in the parentheses are the ratios between the setup time and the time for one solve of
\eqref{eq:lusolve} in double precision.
\label{tab:lapsetupgpu2d3d}}
\begin{subtable}[t]{0.5\textwidth}
\caption{2-D Laplacians with 5-point stencil}
\begin{center}
\renewcommand{\arraystretch}{1.05}
\tabcolsep1.3mm
\begin{tabular}{rrrr}
\multicolumn{1}{c}{LEVR} & \multicolumn{1}{c}{LEVC} & 
\multicolumn{1}{c}{SLFR} & \multicolumn{1}{c}{SLFC} 
\tabularnewline
\hline
34 (1.2) & 27 (1.0) & 31 (2.7) & 26 (3.2)  \tabularnewline
35 (1.0) & 28 (.82) & 32 (2.3) & 27 (2.7)  \tabularnewline
39 (.66) & 32 (.56) & 36 (1.6) & 31 (1.9)  \tabularnewline
56 (.50) & 49 (.45) & 53 (1.4) & 48 (1.8)  \tabularnewline
95 (.43) & 88 (.41) & 92 (1.1) & 87 (1.8)  \tabularnewline
\hline 
\end{tabular}	
\end{center}
\end{subtable}\hfill
\begin{subtable}[t]{0.5\textwidth}
\caption{2-D Laplacians with 9-point stencil}
\begin{center}
\renewcommand{\arraystretch}{1.05}
\tabcolsep1.3mm
\begin{tabular}{rrrr}
\multicolumn{1}{c}{LEVR} & \multicolumn{1}{c}{LEVC} & 
\multicolumn{1}{c}{SLFR} & \multicolumn{1}{c}{SLFC} \tabularnewline
\hline
40 (.97) & 29 (.72) & 37 (2.1) & 29 (2.4) \tabularnewline
43 (.70) & 33 (.54) & 40 (1.6) & 32 (1.8) \tabularnewline
58 (.51) & 47 (.42) & 54 (1.1) & 46 (1.5) \tabularnewline
93 (.42) & 82 (.38) & 90 (.88) & 82 (1.4)  \tabularnewline
167 (.38) & 156 (.36) & 164 (.86) & 156 (1.4)  \tabularnewline
\hline 
\end{tabular}
\end{center}
\end{subtable}

\begin{subtable}[t]{0.5\textwidth}
\caption{3-D Laplacians with 7-point stencil}
\begin{center}
\renewcommand{\arraystretch}{1.05}
\tabcolsep1.3mm
\begin{tabular}{rrrr}
\multicolumn{1}{c}{LEVR} & \multicolumn{1}{c}{LEVC} & 
\multicolumn{1}{c}{SLFR} & \multicolumn{1}{c}{SLFC} \tabularnewline
\hline
39 (4.3) & 22 (2.5) & 34 (2.4) & 21 (2.1)  \tabularnewline
41 (4.7) & 23 (2.8) & 36 (2.7) & 23 (2.6)  \tabularnewline
50 (5.6) & 33 (3.7) & 47 (4.5) & 33 (5.1)  \tabularnewline
62 (4.1) & 46 (3.1) & 57 (5.8) & 45 (6.3)  \tabularnewline
73 (2.5) & 59 (2.1) & 70 (5.0) & 58 (5.7)  \tabularnewline
\hline 
\end{tabular}	
\end{center}
\end{subtable}\hfill
\begin{subtable}[t]{0.5\textwidth}
\caption{3-D Laplacians with 27-point stencil}
\begin{center}
\renewcommand{\arraystretch}{1.05}
\tabcolsep1.3mm
\begin{tabular}{rrrr}
\multicolumn{1}{c}{LEVR} & \multicolumn{1}{c}{LEVC} & 
\multicolumn{1}{c}{SLFR} & \multicolumn{1}{c}{SLFC} \tabularnewline
\hline
93 (7.3)  & 44 (3.4)  & 88 (6.6) & 43 (4.9) \tabularnewline
109 (6.1)  & 61 (3.5) & 106 (8.9) & 61 (7.0) \tabularnewline
140 (4.7) & 92 (3.1) & 134 (8.3) & 92 (7.6) \tabularnewline
166 (2.9) & 115 (2.1) & 162 (5.8) & 115 (5.6)  \tabularnewline
180 (1.6) & 128 (1.2) & 170 (3.3) & 128 (3.6)  \tabularnewline  
\hline 
\end{tabular}
\end{center}
\end{subtable}

\end{table}

Finally, in Table \ref{tab:genmatsetup}, we  present the cost of the setup phases  for the general matrices listed in Table~\ref{tab:gpumatrices}.
In columns 2--5, the numbers outside the parentheses are the time required by the setup phases running on the CPU while the numbers in the parentheses
are the setup time on the GPU. For 11  out of the 12 matrices, running the setup phases on the GPU provided speedups, and for 6 matrices the setup phase
of the solver \texttt{SLFC} was cheaper than that of \texttt{cusparse?\_csrsv2}.

Before closing the experimental results section, we  use the general matrices 
to demonstrate the justification for paying extra cost in setup phases discussed in Section~\ref{sec:intro}, by showing the smallest numbers of the solves, 
denoted by $n_{s}$, needed to make the \texttt{SLFC} solver have a shorter total time, i.e.,
find the smallest $n_s$ such that
\begin{equation} \notag
T^{total}_{\mathtt{SLFC}}= T^{setup}_{\mathtt{SLFC}} + n_s \times T^{solve}_{\mathtt{SLFC}} < T^{setup}_{\mathtt{*}} + n_s \times T^{solve}_{\mathtt{*}}=T^{total}_{*},
\end{equation}
where $T^{setup}_{*}$ and $T^{solve}_{*}$ denote the
setup time and the solve time for a solver respectively.
We will compute the $n_s$'s of \texttt{SLFC} and \texttt{cusparse?\_csrsv2}
which are the two solvers that yielded the best performance  in the solve phases in general for all the test matrices.
We will also compute the $n_s$'s of \texttt{SLFC} and \texttt{GSF} as the \texttt{GSF} solver has an extremely inexpensive setup phase.
These numbers  are given in Table \ref{tab:ns}. 
As shown in the second column of the table, in order to  beat \texttt{cusparse?\_csrsv2} in total time, 
the number of solves required does not need to be very large.
It is often reasonable to assume to have such numbers of solves in the applications 
of iterative solvers.
 The zeros  indicate the cases for which \texttt{SLFC} has a more efficient setup 
 phase and the solve phases of \texttt{SLFC} are also faster than \texttt{cusparse?\_csrsv2}. 
 The numbers for \texttt{GSF}
 shown in the third column are very small since the solve phase performance of this solver
 is generally much worse, in spite of the very cheap setup phase needed by this solver.

\begin{table}[ht]
\caption{
Timings of the setup phases of the SpTrSv algorithms running on the CPU and the GPU 
for the general matrices, measured in milliseconds.
\label{tab:genmatsetup} }
\begin{center}
\renewcommand{\arraystretch}{1.05}
\tabcolsep1.4mm
\begin{tabular}{l|rrrr|rrr}
\multicolumn{1}{c|}{Matrix} & LEVR & LEVC & SLFR & SLFC & CUS1 & CUS2 & GSF \tabularnewline
\hline
af\_shell8  & 76 (88) & 57 (71) & 89 (85) & 57 (70) & 156 & 72 & .43 \tabularnewline
ecology2  & 28 (32) & 39 (26) & 33 (30) & 39 (25) & 82 & 216 & .34 \tabularnewline
webbase1M  & 33 (15) & 32 (11) & 37 (13) & 33 (11) & 59 & 8.0 & 1.0 \tabularnewline
elasticity2D  & 62 (62) & 48 (47) & 72 (59) & 49 (47) & 132 & 265 & .46 \tabularnewline
elasticity3D & 466 (127) & 422 (61) & 525 (120) & 425 (61) & 358 & 52 & 2.2 \tabularnewline 
thermal2 & 55 (37)  & 57 (27) & 62 (34) & 57 (26) & 94 & 11 & .43 \tabularnewline
atmosmodd  & 46 (27) & 38 (17) & 54 (25) & 38 (16) & 80 & 29 & .46 \tabularnewline
StocF-1465  & 103 (92) & 101 (71) & 118 (89) & 102 (71) & 184 & 24 & .81 \tabularnewline
af\_shell10  & 230 (200) & 173 (154) & 268 (195) & 175 (154) & 396 & 372 & 1.2 \tabularnewline
Transport  & 107 (50) & 69 (27) & 125 (46) & 70 (27) & 149 & 37 & .90 \tabularnewline
Bump\_2911   & 664 (354) & 624 (250) & 757 (344) & 631 (251) & 751 & 136 & 3.7 \tabularnewline 
Queen\_4147  & 1872 (726) & 1676 (472) & 2106 (710) & 1690 (469) & 1625 & 253 & 9.3 \tabularnewline 
\hline 
\end{tabular}
\end{center}
\end{table}


\begin{table}[ht]
\caption{
The smallest numbers of solves required by \texttt{SLFC} to have a faster total time (setup time plus
solve time) than \texttt{cusparse?\_csrsv2} and  \texttt{GSF} for the general matrices.
\label{tab:ns} }
\begin{center}
\renewcommand{\arraystretch}{1.05}
\begin{tabular}{l|r|r}
\multicolumn{1}{c|}{Matrix} &  CUS2 & GSF \tabularnewline
\hline
af\_shell8 & 0 & 1 \tabularnewline
ecology2 & 0 & 1 \tabularnewline
webbase1M  &  3 & 9\tabularnewline
elasticity2D & 0 & 1\tabularnewline
elasticity3D & 3 & 1\tabularnewline 
thermal2 & 61 & 3\tabularnewline
atmosmodd & 0 & 1 \tabularnewline
StocF-1465 & 28 & 2  \tabularnewline
af\_shell10 & 0 & 1\tabularnewline
Transport & 0 & 1\tabularnewline
Bump\_2911  & 5 & 2 \tabularnewline
Queen\_4147  & 11 & 1 \tabularnewline 
\hline 
\end{tabular}
\end{center}
\end{table}

\section{Conclusions} \label{sec:conclusion}
This paper considers efficient parallel algorithms for solving sparse triangular linear systems 
on modern many-core processors such as GPUs.
The existing algorithms based on level-scheduling methods are carefully examined and 
new algorithms with self-scheduling schemes are introduced. 
The implementations of these algorithms in CUDA are discussed in great detail.
All the parallel algorithms considered in this paper require a setup phase, where the 
parallelism available in the solve phase is uncovered. 
The justification of paying the extra cost in the setup phase but having
a faster solve phase is provided for the scenarios of several important applications.
The algorithms for performing the setup phases on both CPUs and GPUs were explored.
Experimental results showed that the GPU algorithm can speedup the setup phase for 3-D Laplacian
matrices and the general test matrices considered in this paper.
We remark that there is still room to improve the current algorithm for running the setup phases on GPUs,
which is left as a future endeavor.

Numerical results for structured  problems and general sparse matrices
prove the efficiency of the  solve stages of the proposed algorithms, which can 
outperform the state-of-the-art solvers in cuSPARSE by a factor of up to $2.6$.

\section*{Acknowledgment}
The author acknowledges fruitful discussions with Weifeng Liu on their global-synchronization-free SpTrSv algorithms.

\bibliographystyle{IEEEtran}
\bibliography{local}

\end{document}

%% file: preamble.tex
\def\myrefs#1#2{ 
{\bigskip \noindent
{\Large \bf #2}  
 \list {[\arabic{enumi}]}{\settowidth\labelwidth{[#1]}
 \leftmargin\labelwidth 
 \advance\leftmargin\labelsep
 \usecounter{enumi} }  
 \def\newblock{\hskip .11em plus .33em minus .07em}
 \sloppy\clubpenalty4000\widowpenalty4000
 \sfcode`\.=1000\relax}  }

\def\inv{^{-1}}%
\def\comb#1,#2,{ \left( {#1 \atop #2 } \right)  }%
\def\prodd#1,#2,#3,{ \prod_{\scriptstyle #1 \atop\scriptstyle #2 }^{#3} }%
\def\summ#1,#2,#3,{ \sum_{\scriptstyle #1 \atop\scriptstyle #2 }^{#3} }%

\def\RR{\mathbb{R}}

\newtheorem{algor}{{\sc Algorithm}}[section]

\newtheorem{tabl}{Table}[section]

\def\betab{\begin{tabbing} 
xxxx\=xxxx\=xxx\=xx\=xx\=xx\=xx\=xx\=xx\=xx\=xx\=xx\=xx\= \kill} 
\def\entab{\end{tabbing}\vspace{-0.12in}}

\newcommand{\eq}[1]{\begin{equation}\label{#1}}
\newcommand{\en}{\end{equation}}

\newcommand{\beeq}[1]{\begin{equation}\label{#1}}
\newcommand{\eneq}{\end{equation}}